\documentclass{ieeeaccess}
\usepackage[noadjust]{cite}
\usepackage{amsmath,amssymb,amsfonts}
\usepackage{algorithmic}

\usepackage{textcomp}

\usepackage{xspace}
\usepackage{graphicx}
\usepackage{color}
\usepackage{subcaption}
\definecolor{accessbl}{cmyk}{1, 0.4, 0.3, 0}
\DeclareCaptionFont{bl}{\color{accessbl}}
\usepackage{epsfig,endnotes,paralist,multirow}
\usepackage{wrapfig}
\usepackage{epsfig}
\usepackage[normalem]{ulem}
\usepackage{url}
\usepackage{soul}
\usepackage{multirow}
\usepackage{array}
\usepackage{booktabs}

\def\includeComments{include}
\def\includ{include}
\def\comm[#1]{\ifx\includeComments\includ  \bl \texttt{\textbf{\textit{Note: #1}}} \par \fi}
\def\inlinecomm[#1]{\ifx\includeComments\includ  \textit{Note: #1} \fi}

\newcommand{\nblue}[1]{\textcolor{black}{#1}}

\newcommand{\red}[1]{\textcolor{black}{#1}}

\newcommand{\blue}[1]{\textcolor{black}{#1}}

\newcommand{\nn}{\ensuremath{\phi}\xspace}

\newcommand{\bluee}[1]{\textcolor{black}{#1}}

\newcommand{\para}[1]{\smallskip \noindent {\bf #1}}
\newcommand{\softpara}[1]{\smallskip \noindent \underline{#1}}
\newcommand{\vsp}{\vspace{0.01in}}
\newcommand{\cbb}{\color{black}}
\newcommand{\cbl}{\color{black}}

\def\blackbox{\hfill {\vrule height6pt width6pt depth0pt}}

\newcounter{theorem}
\newtheorem{thm}{Theorem}

\newtheorem{corollary}{Corollary}

\newtheorem{defin}{Definition}
\newtheorem{ex}{Example}
\newtheorem{lem}{Lemma}
\newtheorem{ob}{Observation}

\newenvironment{theorem}{\vsp \begin{thm} \nopagebreak}{{\hfill$\blackbox$} \end{thm} \vsp}
\newenvironment{thm-prf}{\vsp \begin{thm} \nopagebreak}{\end{thm}}

\newenvironment{lem-wo-prf-box}{\vsp \begin{lem} \nopagebreak}{\end{lem}}
\newenvironment{lem-prf}{\vsp \begin{lem} \nopagebreak}{\end{lem}}
\newenvironment{prf}{{\it Proof:} \nopagebreak }{{\hfill$\blackbox$}}
\newenvironment{definition}[1]{\vsp\begin{defin}\begin{rm}(\textsc{#1})} {{\hfill$\Box$} \end{rm}\end{defin} \vsp}

\newenvironment{cor-prf}{\vsp \begin{corollary} \nopagebreak}{\end{corollary}}

\newcounter{packednmbr}

\newcommand{\gsg}{\mbox{\tt GSG}\xspace}

\newcommand{\eps}{\mbox{EP}\xspace}
\newcommand{\epss}{\mbox{EPs}\xspace}
\newcommand{\es}{\mbox{ES}\xspace}

\newcommand{\eat}[1]{}

\let\emph\textit

\newcommand{\php}{\mbox{$p_{ob}$}\xspace}          
\newcommand{\bt}{\mbox{$t_{b}$}\xspace}         
\newcommand{\ft}{\mbox{$t_{f}$}\xspace}         
\newcommand{\bp}{\mbox{$p_{b}$}\xspace}          
\newcommand{\fp}{\mbox{$p_{f}$}\xspace}          
\newcommand{\gt}{\mbox{$t_g$}\xspace}      
\newcommand{\gp}{\mbox{$p_g$}\xspace}       
\newcommand{\ep}{\mbox{$p_e$}\xspace}       
\newcommand{\ct}{\mbox{$t_c$}\xspace}      

\newcommand{\fidl}{\ensuremath{\tau_l}\xspace}
\newcommand{\fidd}{\ensuremath{\tau_d}\xspace}

\newcommand{\fusionTree}{\ensuremath{\mathcal{F}}\xspace}
\newcommand{\links}{\ensuremath{\mathcal{E}}\xspace}

\newcommand{\ket}[1]{\ensuremath{|#1\rangle}}

\newcommand{\ghzt}{\mbox{\tt 3-Star}\xspace}

\newcommand{\ghz}{\mbox{GHZ}\xspace}
\newcommand{\central}{\mbox{\tt Central}\xspace}

\newcommand{\idj}{i\cdotp\cdotp j}
\newcommand{\idk}{i\cdotp\cdotp k}
\newcommand{\idi}{i\cdotp\cdotp i}
\newcommand{\jdk}{j\cdotp\cdotp k}
\newcommand{\rds}{r\cdotp\cdotp s}
\newcommand{\sdt}{s\cdotp\cdotp t}
\newcommand{\rdt}{r\cdotp\cdotp t}

\newcommand{\oslp}{\mbox{\tt One-Stage}\xspace}
\newcommand{\tslp}{\mbox{\tt Two-Stage}\xspace}
\newcommand{\llp}{\mbox{\tt Left-Sided}\xspace}
\newcommand{\rlp}{\mbox{\tt Right-Sided}\xspace}
\newcommand{\distance}{\mbox{\tt Distance-Based}\xspace}

\newcommand{\mpfont}{\scriptsize}

\ifx\noeditingmarks\undefined
    \newcommand{\MPworker}[2]{{\color{#1}\vrule\vrule}{\marginpar{\color{#1}\mpfont #2}}}
\else
    \newcommand{\MPworker}[2]{}
\fi

\def\BibTeX{{\rm B\kern-.05em{\sc i\kern-.025em b}\kern-.08em
    T\kern-.1667em\lower.7ex\hbox{E}\kern-.125emX}}
\begin{document}
\history{Date of publication xxxx 00, 0000, date of current version xxxx 00, 0000.}
\doi{10.1109/TQE.2020.DOI}

\title{Optimized Distribution of Entanglement Graph States in Quantum Networks}
\author{
    \uppercase{Xiaojie Fan}\authorrefmark{1},
    \uppercase{Caitao Zhan}\authorrefmark{2},
    \uppercase{Himanshu Gupta}\authorrefmark{1}, and
    \uppercase{C.\ R.\ Ramakrishnan}\authorrefmark{1}
}
\address[1]{Department of Computer Science, Stony Brook University, Stony Brook, NY 11790, USA}
\address[2]{Argonne National Laboratory, Lemont, IL 60439, USA}

\tfootnote{This work was supported by NSF awards FET-2106447 and CNS-2128187.}

\markboth
{Fan \headeretal: Optimized Distribution of Entanglement Graph States in Quantum Networks}
{Fan \headeretal: Optimized Distribution of Entanglement Graph States in Quantum Networks}

\corresp{Corresponding author: Xiaojie Fan (email: xiffan@cs.stonybrook.edu).}

\begin{abstract}
Building large-scale quantum computers, essential to demonstrating quantum advantage, is a key challenge. Quantum Networks (QNs) can help address this challenge by enabling the construction of large, robust, and more capable quantum computing platforms by connecting smaller quantum computers. Moreover, unlike classical systems, QNs can enable fully secured long-distance communication. Thus, quantum networks lie at the heart of the success of future quantum information technologies. 
In quantum networks, multipartite entangled states distributed over the network help implement and support many quantum network applications for 
communications, sensing, and computing.
Our work focuses on developing optimal techniques to generate 
and distribute multipartite entanglement states efficiently.

Prior works on generating general multipartite entanglement states have 
focused on the objective of minimizing the number of maximally 
entangled pairs (\epss) while ignoring the heterogeneity of the 
network nodes and links
as well as the stochastic nature of underlying processes.
In this work, we develop a hypergraph-based linear programming 
framework that delivers\textbf{}
optimal (under certain assumptions) generation 
schemes for general multipartite entanglement represented by graph states, 
under the
network resources, decoherence, and fidelity constraints, while
considering the stochasticity of the underlying processes.
We illustrate our technique by developing generation schemes 
for the special cases of path and tree graph states, and discuss
optimized generation schemes for more general classes of graph 
states.
Using extensive simulations over a quantum network simulator (NetSquid), 
we demonstrate the effectiveness of our developed techniques and show that they
outperform prior known schemes by up to orders of magnitude.

\end{abstract}


\titlepgskip=-15pt

\maketitle

\section{\bf Introduction}
\label{sec:intro}

Quantum networks (QNs) enable the construction of large-scale and robust quantum computing platforms by connecting smaller QCs~\cite{qn}. 
QNs also enable various important applications~\cite{qsn-qce-23,qsn,qkd,komar2014quantumclock,qsn-phyA-23,secure,qsn-acm-24,byzantine,dqc-disc-q21,zhan-thesis-24}, but to implement and support many of these applications, we need to create and distribute entangled states efficiently~\cite{swapping-tqe-22,zhan-acp-qcnc25,predist-qce-22,sundaram2024optimized,fan2025qcnc,gyongyosi2019opportunistic}. 
Recent works have addressed the generation of entanglement
states but in limited settings, e.g., bipartite and GHZ states,
or graph states with a simplistic optimization objective.
In this paper, we consider the generation and distribution of specialized graph
states over quantum networks, with minimal generation latency, taking into consideration 
the stochastic nature of the underlying generation process.

\para{Graph States and Their Applications.}
Graph states are multipartite entangled states where a graph over the qubits 
specifies the entanglement structure between qubits. 
Owing to their highly entangled nature, graph states find applications
in various quantum information processing domains, such as measurement-based
quantum computing, quantum error correction, quantum secret sharing, and quantum metrology. In particular, path/cycle graph states are used as a primary resource state of fusion-based quantum computing~\cite{bartolucci2023fusion}, and tree graph states find usage in counterfactual error correction~\cite{error_correction01}, photonic measurement-based quantum computing, and fusion-based quantum computing~\cite{bartolucci2023fusion}. The star graph state, which is a special case of a tree graph state, is equivalent to a GHZ state---which has many applications, including error correction~\cite{error_correction01}, quantum secret sharing~\cite{ghz_secret99},
quantum metrology~\cite{metrology14},  clock synchronization~\cite{komar2014quantumclock}, etc.
Therefore, developing efficient generation schemes to distribute graph 
states in a QN is of great significance. 
Our work focuses on developing optimal generation schemes for 
general classes of graph states.

\para{Prior Work and Our Approach.}
There have been recent works~\cite{star_expansion19,maxflowQCE21,korean23} that have 
addressed the problem of efficient generation and
distribution of general graph state entanglements in a quantum network.
These works, however, have focused on the simplistic optimization objective of 
minimizing the \emph{number} of maximally entangled pairs (e-bits or \epss) consumed; 
in particular, they implicitly ignore the stochastic nature of the underlying processes. 
Even a true {\it count} of \epss consumed should consider the stochastic nature of operations 
(e.g., fusion) involved, particularly since they can
have a relatively low probability of success. 
Moreover, some \epss may take significantly longer to generate than others due to the 
heterogeneity of the network. Thus, the number of \epss is too simplistic a performance metric.

In this work, we consider the generation and distribution of classes of 
graph states to maximize the expected generation rate under given network resource and fidelity constraints while considering the stochastic nature of underlying processes and network heterogeneity. This is in the same vein as the recent works on the 
generation of \epss~\cite{swapping-tqe-22, sigcomm20, delft-lp, caleffi} and 
GHZ states~\cite{ghz-qce-23} in quantum networks.
In particular, our goal is to develop {\it provably optimal} generation schemes.
We develop a 
framework---based on a hypergraph representation of the 
intermediate graph states and fusion operations---that delivers optimal (under reasonable assumptions) generation schemes under network and fidelity constraints.
We illustrate our framework by developing multiple generation schemes for
the path and tree graph states, and discuss generalizations to other classes of graphs.
In essence, our proposed schemes use fusion operations to build larger graph states from smaller ones progressively and discover the optimal level-based structure (that represents 
the generation process, i.e., sequence and order of fusion operations over intermediate graph states) by using an appropriate linear programming formulation.

\para{Our Contributions.}
In the above context, we make the following contributions.
\begin{enumerate}
    \item 
    We develop a framework for developing optimal schemes for generating graph states in quantum networks under network resource and fidelity constraints, considering the stochastic nature of the fusion operations.
    
    \item Specifically, for path graph states, we design a polynomial-time generation scheme that is provably optimal under reasonable assumptions. In addition, we also develop an optimal two-stage generation scheme that is computationally more efficient, based on restricting the intermediate graph states created.

    \item Similarly, for tree graph states, we design two generation schemes that are optimal under the restriction on the intermediate states and fusion operations used.
    
    \item We show the versatility of our developed scheme by discussing and illustrating its application for other classes of graph states, e.g., grid graphs, bipartite graphs, and complete graphs. We also generalize our scheme to generate multiple graph states concurrently. 

    \item Using extensive evaluations over the NetSquid simulator, we demonstrate the effectiveness of our developed techniques and show that they outperform prior work by up to orders of magnitude.
\end{enumerate}               
\section{\bf Background}
\label{sec:background}

\para{Quantum Network (QN), Nodes, Links, and Communication.} 
A quantum network (QN) is a network of 
quantum computers (QCs), and is represented as a connected undirected graph with vertices 
as QCs and edges representing the (quantum and classical) direct 
communication links. We use {\it network nodes} to refer
to the vertices (QCs) and {\it links} to refer to the edges,
in the QN graph.
We discuss a detailed network model in \S\ref{sec:problem}. 
Since direct transmission of quantum data is subject to unrecoverable errors, especially
over long distances, we use teleportation to transfer quantum information reliably across
nodes in a QN. Teleportation requires that a maximally-entangled pair (\eps) 
be already established over communicated nodes.

\begin{figure}
    \centering
    \includegraphics[width=0.46\textwidth]{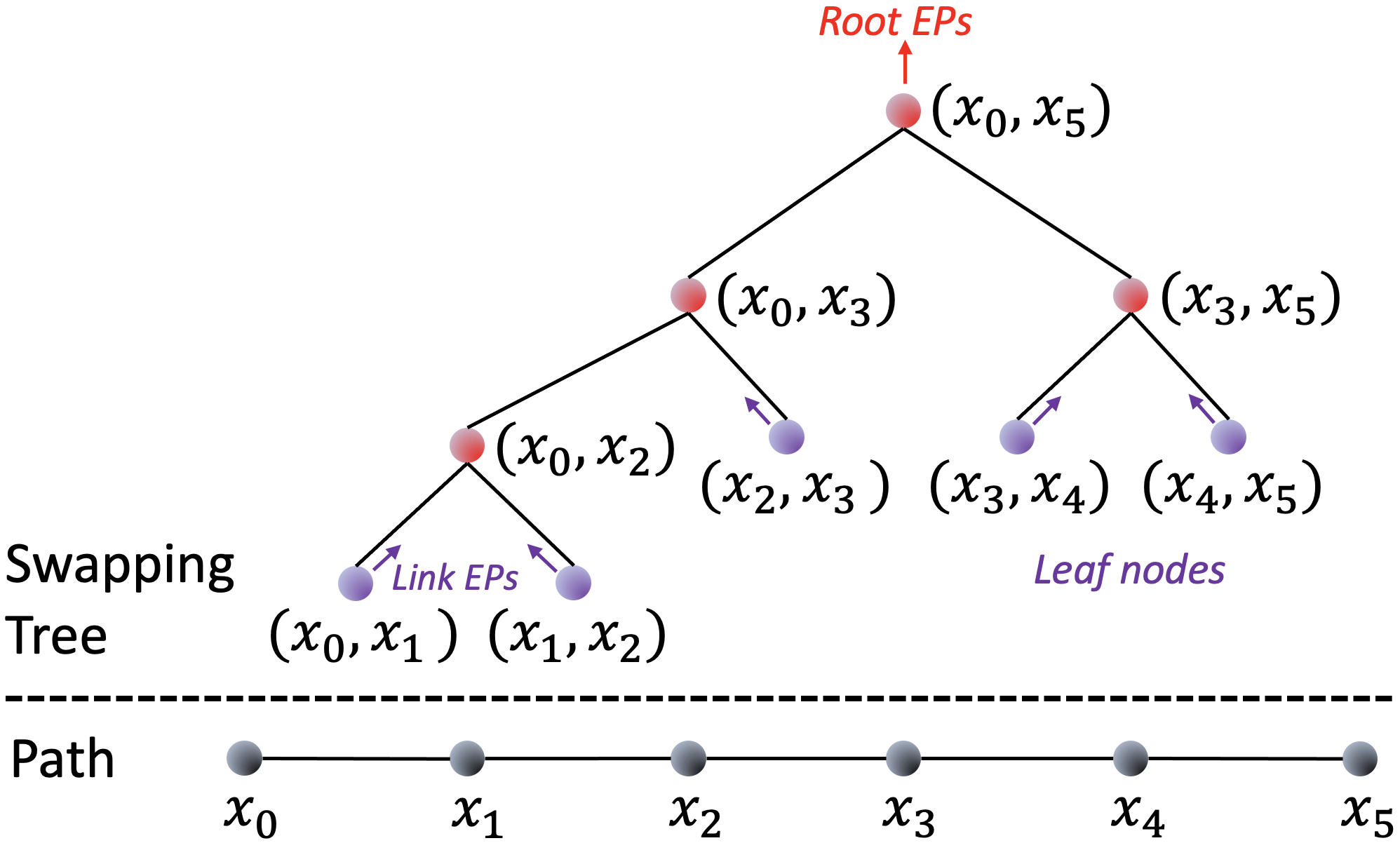}
    \vspace{0.05in}
    \caption{A swapping tree over a path.  The leaves of the tree are the link \epss, which are being generated continuously. Here, the notation $(x_i, x_j)$ represents an \eps over two qubits residing in the network nodes $x_i$ and $x_j$.}
\vspace*{-0.2in}
\label{fig:swapping_tree}
\end{figure}

\para{Generation of 
Remote \epss using Swapping Trees.} 
An efficient way to generate an \eps over a pair of 
remote network nodes $(s, d)$ using \epss over 
network links (i.e., edges) is to: 
(i) create a path $P$ in the network graph from $s$ to $d$ 
with \epss over each of the paths' edges, and (ii) perform
a series of entanglement swaps (ES) over these \epss. 
The series of ES operations over $P$
can be performed in any arbitrary order, 
but this order of ES operations affects the latency
incurred in generating the \eps over $(s,d)$.
One way to represent the ``order'' in which the
ES operations are executed---is a complete binary
tree over the link \epss as leaves, called a 
{\it swapping tree}~\cite{swapping-tqe-22}. 
See Fig.~\ref{fig:swapping_tree} (from~\cite{swapping-tqe-22}).
The stochastic nature of ES operations entails that 
generation of an \eps over a remote pair of nodes 
using a swapping tree may incur
significant latency, called the {\it generation latency} (inverse
of generation rate). Generation latency is largely due to the latency 
incurred in (i) generating the link \epss, 
and (ii) a generated \eps $(x_i, x_j)$ {\it waiting} 
for its ''sibling'' \eps $(x_j, x_k)$ to be generated before an 
ES operation can be performed over them to generate an \eps over 
$(x_i, x_k)$.

\begin{figure}[h]
    \vspace{-0.1in}
    \centering
    \includegraphics[width=0.5\textwidth]{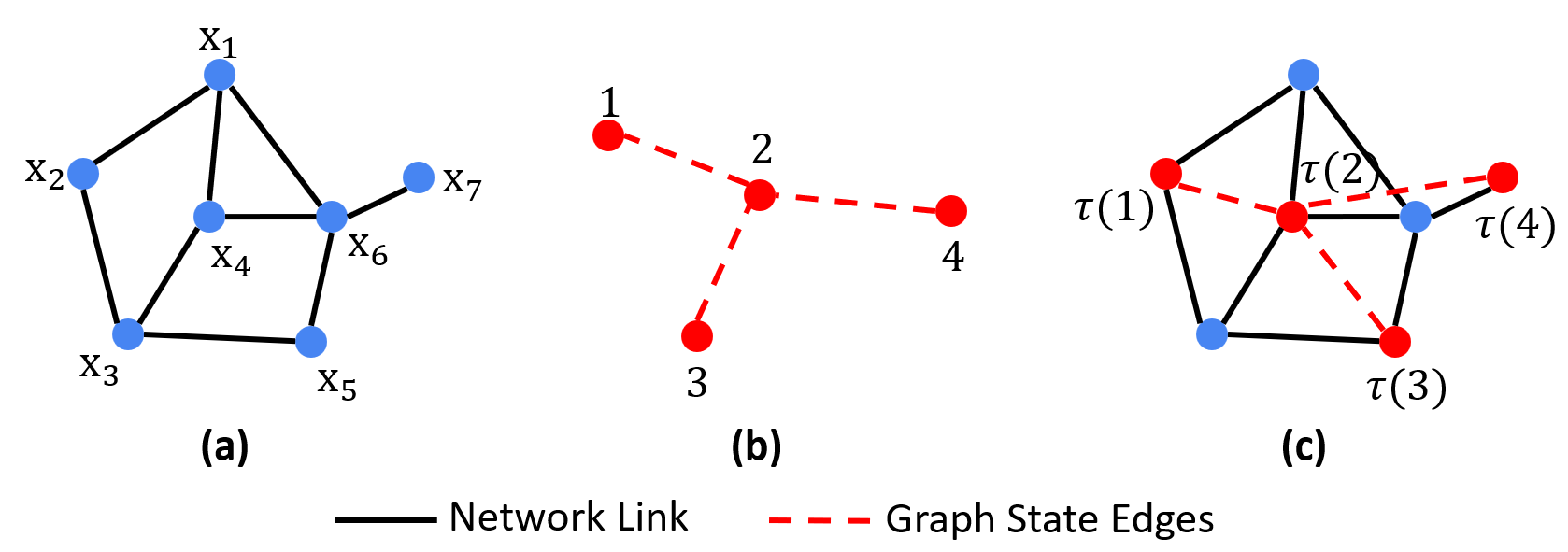}
    \caption{(a) A quantum network with 7 nodes $\{x_1, x_2, \ldots, x_7\}$, (b) Graph State $G$ with 4 vertices named 1 to 4, and (c) Distributed Graph State (in red) for the graph state $G$ with $\tau(1) = x_2, \tau(2) = x_4, \tau(3)=x_5, \tau(4)=x_7$.}
    \vspace{-0.15in}
    \label{fig:graphs}
\end{figure}

\subsection{\bf Graph States: Multipartite Entanglements}
\label{sec:back-multi}


\para{Distributed Graph States; Link States.}
A {\it graph state} is a multipartite quantum state $\ket{G}$ which is described by a graph $G$, where the vertices of G
correspond to the qubits of $\ket{G}$. 
Formally, a graph state $\ket{G}$  is given as
$$\ket{G} = \Pi_{(u,v) \in E(G)} C_Z^{(u,v)} \otimes_{v \in V(G)} \ket{+}_v,$$
where $C_Z^{(u,v)}$ is the controlled-Z (CZ) gate over the qubits $u$ and $v$.
We use the term {\it distributed} graph state to mean a graph state $G$ along with its (target or current) distribution over the given quantum network; this distribution is
represented by a function $\tau: V(G) \mapsto V(Q)$ of graph state's vertices $V(G)$ to the 
nodes $V(Q)$ in the given quantum network $Q$. 
See Fig.~\ref{fig:graphs}.\footnote{We generally use $x_i$'s and $y$ for network nodes, and numbers 1, 2, etc.\ for graph state's vertices.}
For brevity, when clear from the context, we just use {\it states} to 
refer to {\it distributed graph states.}
Also, we use the term  {\it {\bf link states}} to refer to the
single-edge graph states distributed over the network links; these link
states are locally equivalent to the link-EPs generated by
the adjacent nodes.

\para{Generation\footnote{Throughout the paper, by {\it generation} of states, we implicitly mean generation and distribution of created states.} of Graph States via Fusion Trees.}
We need to fuse smaller graph states and/or modify graph states to generate general graph states. In general, starting with link states, 
we want to generate graph states using only {\it local} quantum operations (i.e., gates with operands in a single node).
Similar to swapping trees used to describe the generation of EPs, we can use {\it fusion trees} to describe the generation of graph states in a QN using local fusion operations. Each node in a fusion tree would represent a distributed graph state.
Such fusion trees have been used in prior works---e.g., for generating and distributing GHZ states~\cite{ghz-qce-23}.

\begin{figure}[h]
    \vspace{-0.1in}
    \centering
    \includegraphics[width=0.5\textwidth]{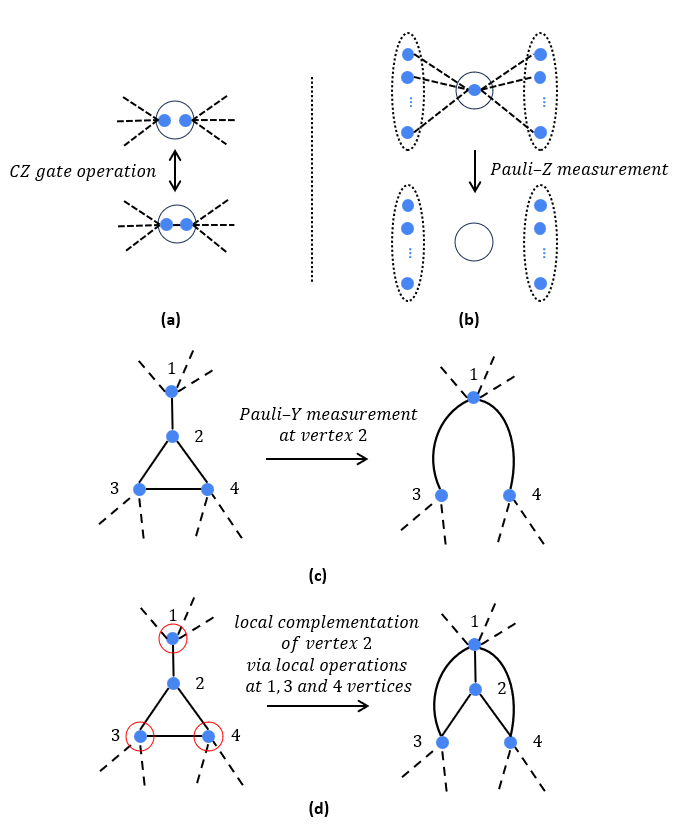}
    \caption{Local operations used in our fusion operations.}
    \vspace{-0.15in}
    \label{fig:ops}
\end{figure}

\softpara{Fusion Operations.} Local operations within a fusion are generally restricted to single-qubit Clifford operations, local CZ gates, or Pauli measurements. In our context, we only use the following  operations or measurements within a local fusion operation (see Fig.~\ref{fig:ops}):

\begin{enumerate}[(a)]
\item 
Create or remove an edge (in the graph state) by doing a CZ operation over two vertices (of the graph state). This operation is local when the qubits corresponding to the vertices are available in a single node.

\item 
Pauli-Z measurement over a qubit/vertex $q$ results in $q$'s deletion.

\item 
Pauli-Y measurement over a qubit/vertex $q$ results in a local complementation of vertex $q$'s neighborhood and then $q$'s deletion.

\item 
Also, one can effectuate local complementation of any vertex $q$ by doing appropriate single-qubit Clifford Operations at its neighbors.
\end{enumerate}

\section{\bf Model, Problem, and Related Works}
\label{sec:problem}

In this section, we discuss our network model, 
formulate the problem addressed, and discuss 
related work. 

\para{Network Model.} 
We denote a quantum network (QN) $Q$ with $V(Q)$ 
denoting the set of nodes. Adjacent nodes signify nodes
connected by a communication link. 
Our network model is similar to the one used in 
some of the recent works~\cite{swapping-tqe-22,ghz-qce-23} on
efficient generation of \epss and GHZ states.
In particular, each node has an atom-photon \eps generator with generation 
latency (\gt) and probability of success (\gp); 
\nblue{the atom-photo generation latency refers to the time interval between consecutive attempts by a node to excite an atom for the purpose of generating an atom-photon entangled pair,} 
which implicitly includes other latencies incurred in link \eps generation 
viz.\ photon transmission, optical-BSM, and classical acknowledgment.
A node's atom-photon generation capacity/rate 
is its aggregate capacity and may be split across its incident links.
Each network link $e=(A,B)$ is used to generate link-EPs, \blue{which is locally equivalent to single-edge graph states, }
using an optical BSM device located in the middle. 
The optical-BSM  has 
a certain probability of success (\php); and each half-link 
(from $A$ or $B$) to the device has a probability of transmission success (\ep) that decreases exponentially with the link distance.
To facilitate atom-atom \es and fusion operations, 
each network node is also equipped 
with an atomic-BSM device 
with appropriate latency
and probability of success. 
There is an independent classical network with a transmission latency of \ct;
we assume classical transmission always succeeds.

\begin{wrapfigure}{r}{1.6in}
    \centering
    \includegraphics[width=0.23\textwidth]{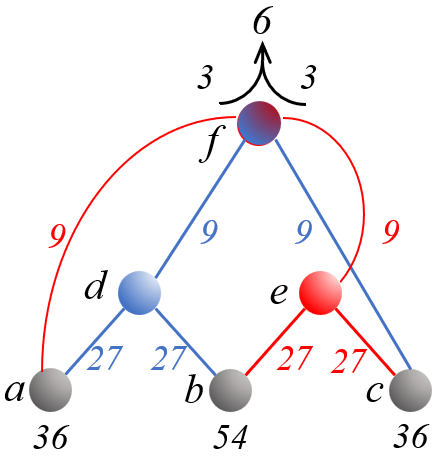}
    \caption{Level-based structure. The above structure is an ``aggregation'' of two fusion trees. The leaf node $a$'s generation rate of 36 units is ``split'' into 9 and 27 for the two different (red and blue) fusion operations. The root node represents the final/target graph state formed in two different ways---for a total generation rate of 6 (3 from each fusion operation). We assume that a parent's generation rate is 1/3 of the rate of its children/operands (which are equal).}
\label{fig:level}
\end{wrapfigure}

\para{Level-Based Fusion Structure.}
To maximize the generation rate of a graph state,
{\it multiple concurrent} fusion trees may be required to use 
all available network resources.
Since the number of such trees can be exponential, 
we use a novel ``aggregated'' structure that aggregates multiple 
fusion trees into one structure; we refer to this 
as a {\it level-based} structure, as it is composed of multiple 
levels---with each level consisting of distributed graph states (as vertices) 
created by fusing states from the previous levels. See Fig.~\ref{fig:level}. \blue{(Similar multi-level structure is used in~\cite{dai2020optimal} for 
generation of EPs.)} 
The bottom level consists of link states, and each non-leaf state 
$S$ is formed by a fusion of pairs of states in the previous layers; however, there may be several such pairs of states that fuse to create \red{$S$} (in different ways). 
Each state $S$ may also have multiple ``parents'' (unlike in a tree), i.e., 
a state $S$ may be used to create several states in the next layer; in such a case, the generation rate of $S$ is ``split'' across these fusions.
Due to the fusions from previous layers, each vertex/state has a resulting generation rate, estimated as discussed below.

\blue{In the level-based structure shown in Fig.~\ref{fig:level}, node $f$ represents the target graph state we aim to generate, while nodes $a$, $b$, and $c$ correspond to single-edge states. Nodes $d$ and $e$ belong to two fusion trees within the same quantum network, both contributing to the generation of $f$. The two fusion trees are represented by blue and red edges, respectively.  
The leaf node $a$ has a generation rate of 36 units, which is "split" into 9 and 27 units for the two different fusion operations in the red and blue fusion trees. Each fusion tree individually contributes to generating $f$ with an effective generation rate of 3 units.}

\para{Graph State Generation Latency/Rate.}
The expression for estimating the generation rate (or latency) of a state due to a fusion operation in our level-based structure is fundamentally the same as that used in fusion/swapping trees in prior works~\cite{swapping-tqe-22,ghz-qce-23}. 
Consider a simple case of a non-leaf node $t$ with 
two children $t_l$ and $t_r$ which are fused to generate $t$. 
If the generation events of the children states
$t_l$ and $t_r$ are Poisson distributed and thus generation 
latencies are exponentially distributed,
then, the generation latency of the graph state corresponding 
to $t$ can be estimated as (see~\cite{swapping-tqe-22}):
\begin{equation}
L_t = (\frac{3}{2} \max(L_l, L_r) + \ft + \ct)/\fp, \label{eqn:tree-rate}
\end{equation}
where 
$L_l$ and $L_r$ are the generation latencies of the graph states
corresponding to the children $t_l$ and $t_r$, 
\ft and \fp are the latency
and probability of success of the swapping/fusion operation, and \ct is
the classical transmission latency \blue{which is proportional to 
the physical distance.}
The generation rate $G_t$ can thus also 
be estimated as $G_t = 2/3 \min(G_l, G_r)$, 
where $G_t$, $G_l$, and $G_r$ are the generation rates of the nodes.
For a state $t$ generated from multiple pairs of children, we take the sum of the generation rates due to each pair. 
The generation rate of the leaf vertices (link states) in a level-based structure is given by the generation rate of the \eps at the network link.
To estimate the generation rate of other states in the structure,  
we apply the
above equation iteratively (for this, we implicitly assume that the resulting 
latencies also have an exponential distribution).

\subsection{\bf Problem Formulation}
\label{sec:formulation}
In this section, we formulate the problem of efficiently generating 
distributed graph states over a quantum network. Informally, the problem
is to generate the level-based structure with a maximum generation output rate, given the constraints of the nodes' link-EP generation capacity.

\para{Graph State Generation (\gsg) Problem.} 
Given a quantum network $Q$, a  
graph state $G$ along with its distribution $\tau: V(G) \mapsto V(Q)$, 
the \gsg problem is to determine a level-based structure \fusionTree that 
generates the giving distributed graph state with the optimal (highest) 
generation rate under the following node constraint. (For clarity of presentation, we consider fidelity and decoherence constraints later in
\S\ref{sec:gen}.)
We refer to the given $G$ as the {\it target} graph state, and the network nodes $\tau(i)$, for $i \in [1,n]$, as the {\it terminal} nodes.

\softpara{Node Constraints.}
For each network node, the aggregate resources used by \fusionTree is less than the available resources. More formally, consider a level-based structure \fusionTree. Let $\links$ be the set of all network links, and $\links(i) \subseteq \links$ as the set of links
incident on node $i$. For each link $e \in \links$, let $R(e)$ be the total generation rate of $e$ in \fusionTree. 
Then, the node capacity constraint is formulated as follows.
\begin{eqnarray}
1/\gt &\geq& \sum_{e \in \links(i)} R(e)/(\gp^2\ep^2\php) \ \ \ \  \forall i \in \nblue{V(Q)}. \label{eqn:qnr-1}
\end{eqnarray}
The above comes from the fact that to 
generate an edge graph state over $e$, each end-node of $e$ needs
to generate $1/(\gp^2 \ep^2 \php)$ photons successfully, 
and that $1/\gt$ is a 
node's total generation capacity. 
\eat{Also, the \underline{memory constraint} is that for any node $i$, the 
memory available in $i$ should be more than $|E(i)|$.}



\begin{figure}[h]
     \includegraphics[width=0.5\textwidth]{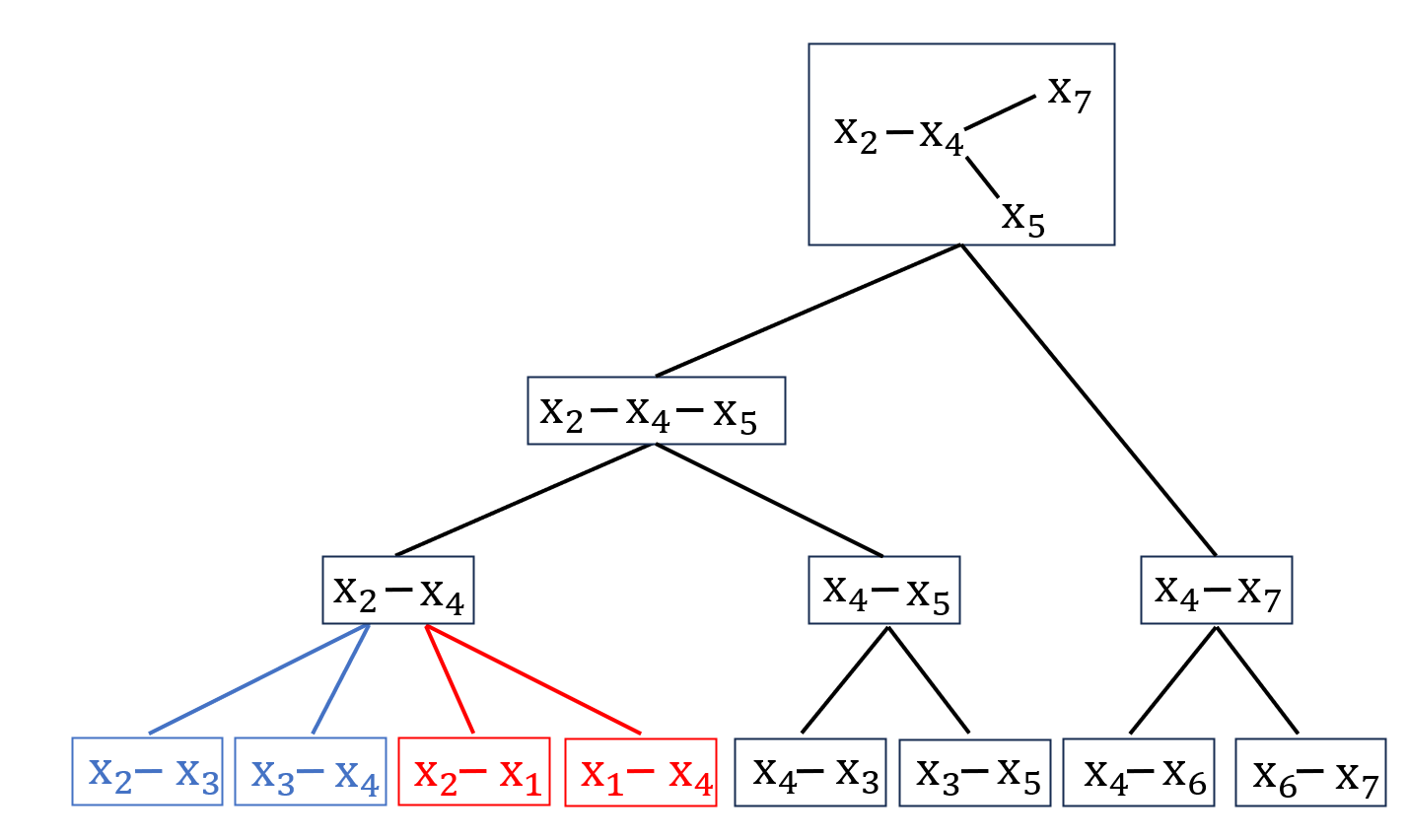}
    \caption{A potential solution (not necessarily optimal), a level-based structure, for the network graph and distributed graph state in Fig.~\ref{fig:graphs}. The distributed graph state corresponding to a node in the structure is represented by the actual graph and its distribution (e.g., $x_2$---$x_4$ represents an edge graph state distributed over nodes $x_2$ and $x_4)$.}
    \label{fig:gsg-example}
\end{figure}

\para{\gsg Example.} Consider the \gsg instance shown in Fig.~\ref{fig:graphs}. For this instance, one possible solution---a level-based structure not necessarily optimal---is shown in Fig.~\ref{fig:gsg-example}. This structure depicts two ways (shown in blue and red) of generating the distributed graph state $x_2$---$x_4$; apart from this, the structure is essentially a fusion tree generating the desired (distributed) target graph state from the link states.

\subsection{\bf Related Works}
\label{sec:related}

There has been recent interest in developing schemes for 
generating  graph states in a quantum network. 
Most of these works have focused on minimizing the {\it number} of link \epss consumed in generating the graph states. We discuss these below, 
categorized by {\it Centralized} and {\it Distributed} schemes.
To the best of our knowledge, there has been no prior work 
on efficient generation of arbitrary graph states (or broad classes 
of graphs) that optimize the generation rates while taking into 
considering the stochastic nature of the underlying 
processes; perhaps, the only exception is~\cite{ghz-qce-23} which 
considers the generation of GHZ states (we discuss this below, in the last paragraph of this subsection). 

\softpara{Centralized Schemes.}
In a centralized generation scheme, an appropriately chosen central node first creates the target graph state locally and then teleports qubits to the 
 terminal nodes using \epss. 
In particular,~\cite{maxflowQCE21} proposes a max-flow-based approach to minimize the number of link-\epss consumed in generating a graph state using such a scheme.
They represent the teleportation routes as multi-path flows and use a network flow
approach to maximize the total generation rate. 
The network-flow approach allows the representation of network resource 
constraints but ignores the stochastic aspect of the teleportation (or entanglement-swapping) process, which fundamentally requires considering the {\it length} of the teleportation paths (ignored in the network-flow representation).

\softpara{Distributed Schemes.}
In a distributed generation scheme, the target graph state is generated in a distributed manner (perhaps by iteratively merging smaller graph 
states)---as in the schemes discussed in this paper. 
In~\cite{star_expansion19}, the authors propose a \textit{star expansion} operation/sub-protocol
to fuse \epss, and use the operation iteratively to generate a 
target \ghz (equivalent to a star graph) state. Then, using a succession of such
star graphs, they create a complete graph
state with appropriate edges ``decorated''---which are removed to yield the target graph state. Their optimization objective is the minimization of the
number of \epss consumed, and more importantly, for sparse graph states, their scheme can be very wasteful. 
In a more recent work,~\cite{korean23} presents a graph-theoretic strategy to 
optimize the fusion-based generation of arbitrary graph states effectively; their strategy comprises three stages: simplifying the graph state, building a fusion
tree/network, and determining the order of fusions. They use 3-qubit GHZ states
as the basic resource and optimize the number of these states used. They do not
discuss techniques to generate and distribute graph states {\it over a quantum network}; nevertheless, we believe theirs is the most promising approach among existing works 
for generating arbitrary graph states in a quantum network.
Thus, we adapt/extend their scheme for distributing graph states in a quantum network 
and compare it to our schemes in~\ref{sec:eval}. 

\softpara{Generating EPs and GHZ States; Our Work.}
Finally, there have been works on the generation and distribution of 
specialized graph states, e.g.,~\epss~\cite{swapping-tqe-22,sigcomm20,delft-lp,dai2020optimal} and 
GHZ states~\cite{ghz-qce-23,dp_ghz_from_bell20,bugalho2021distributing}. 
Our work on generating general graph states uses a similar network model and optimization 
formulation as~\cite{swapping-tqe-22,ghz-qce-23}, but has different objectives and thus uses different techniques.
In particular,~\cite{swapping-tqe-22}
designs a dynamic programming approach to construct optimal swapping trees to generate remote \epss, and~\cite{ghz-qce-23} develops heuristics to construct fusion trees
to generate GHZ states. 
Instead, our objective is to develop
a general framework for the optimal generation of general classes of graph states; in particular, we develop a hypergraph-based framework 
to construct optimal level-based structures (instead of trees) by
determining an optimal hyperflow in hypergraphs.

\section{\bf High-Level Approach}
\label{sec:high-level}

Here, we discuss our overall approach to optimally solving the \gsg problem using linear
programming (LP). In the following sections, we will apply our technique to two special cases of graph states: paths and trees. In \S\ref{sec:gen}, we briefly 
also discuss other classes of graph states to demonstrate the versatility of our
approach.

\begin{figure}[h]
    \centering
    \includegraphics[width=0.5\textwidth]{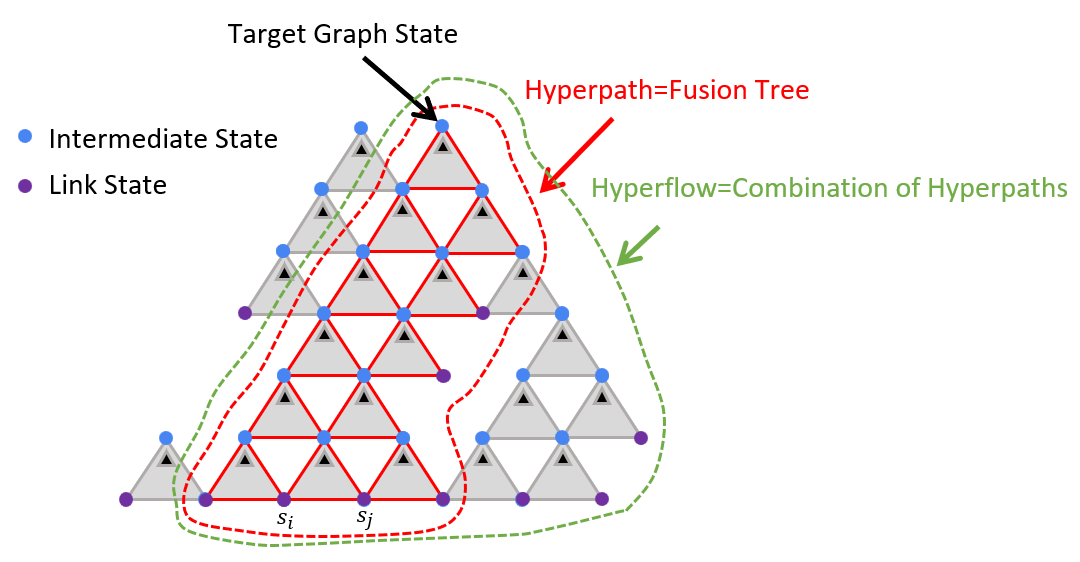}
    \caption{A hypergraph over potential intermediate states. Here, each hyperedge represents a fusion operation and a hyperpath represents a potential fusion tree. A hyperflow incorporates multiple hyperpaths, akin to a network flow in a simple graph, and represents a level-based structure.}
    \label{fig:basic}
\end{figure}
\para{Basic Idea.} Given a quantum network and a target graph state, 
we create a {\it hypergraph} that has embedded in it all possible
level-based structures. In this envisioned hypergraph (see Fig.~\ref{fig:basic}): 
(i) each vertex is a potential intermediate distributed graph state, (ii) each hyperedge ($\{s_1, s_2\}, s_3$), for vertices $s_1, s_2, s_3$,
is a fusion operation that fuses graph states $s_1$ and $s_2$ to create $s_3$,
(iii) a ``hyperpath'' is a potential fusion tree, (iv) and a hyperflow  is a level-based structure; here, a hyperflow is basically a ``combination'' of hyperpaths similar to a network flow in a simple graph being a ``combination'' of simple paths.
To determine the optimal hyperflow (and thus, an optimal level-based structure), we assign flow variables representing generation rates to the hyperedges and create an LP with linear constraints corresponding to network resource constraints, flow conservation, and fusion success probability.
\blue{This essentially transforms the \gsg problem into a max-flow problem on the above hypergraph, where we seek to maximize the flow from the $start$ node, which is used to allocate generation rates to single-edge graph states, to the $term$ node, which aggregates the generation of all the target graph states.
A set of linear equations describes the relationship between the incoming and outgoing flow (generation rate) at each hypervertex, effectively modeling the allowed fusion operations and the corresponding loss in generation rate due to the operation.} A similar LP approach has been used
as a benchmark in our earlier 
work~\cite{swapping-tqe-22} for generating EPs.

\para{Key Challenge.} In general, {\it any} distributed graph state in a given network 
can be considered as an intermediate state in the process of generating a given target graph state;
thus, the number of potential intermediate states is exponential ($O(4^n)$) in the number $n$ of network nodes. 
However, only certain types of intermediate state are likely to be useful/relevant in the generation of a given target state; e.g., to generate a single-edge graph
state, it seems reasonable to consider only 
single-edge graph states as intermediate states (as in the generation of
remote EPs via entanglement swapping, which generates only EPs as 
intermediate states; note that EPs are locally equivalent to single-edge states).
Thus, the key challenge in using the above approach is to determine an appropriate set of
intermediate states such that the resulting LP over the corresponding hypergraph is computationally feasible and delivers a ``good'' solution.\footnote{These good solutions can also be shown to be optimal (as shown in Theorem~\ref{thm:path-one}), under appropriate assumptions.} In particular, we also consider the below two-stage approach to minimize the number
of intermediate states considered.

\softpara{Two-Stage LP Approaches.}
One strategy we consider to minimize the number of intermediate states considered
is to generate the target graph state in two stages: (i) Generate  
single-edge graph states for each edge in the target graph state $G$;
(ii) Use these edge graph states to iteratively generate appropriate  
intermediate states and eventually the target graph state; in this second
stage, only the terminal nodes are involved. We discuss such approaches for
path and tree graph states in the following sections. 
\section{\bf Generating Path Graph States}
\label{sec:fusion_retain}

In this section, we design algorithms to generate distributed path graph states based on the high-level approach described 
above.\footnote{\blue{A path graph state is equivalent to a 1D cluster state in quantum networks}} 
\blue{To correspond with the {\tt Two-Stage LP} mentioned later, we refer to this method as {\tt One-Stage LP}. }
We recall the standard hypergraph notion. 

\begin{definition}{Hypergraph}
\label{defn:hypergraph}
\rm
  A directed hypergraph $H = (V(H), E(H))$  has a set of vertices $V(H)$
  and a set of (directed) \emph{hyperedges} $E(H)$, where each hyperedge $e$ is a
  pair $(t(e), h)$ with the {\it tail} $t(e) \subset V(H)$ and the {\it head} $h \in (V(H) - t(e))$.\footnote{In general hypergraphs, $h$ can also be a subset of $V(H)$, but in our context, $h$ is just a single vertex. Also, in our schemes, $|t(e)|$ is just 1 or 2.} 
\end{definition}





\subsection{\textbf{Optimal Generation of Path Graph States}}
\label{sec:opt-path}

Consider a \gsg problem instance, wherein the target graph state $G$ is a path graph 
 $(1, 2, 3 \ldots, n)$ with edges $(i, i+1)$ for all $1 \leq i < n$ and the
target distribution represented by $\tau: V(G) \mapsto V(Q)$.

\para{Basic Idea.}
For the path state, we 
\bluee{hypothesize}
that the type of intermediate states that can potentially be useful in generating and distributing the path state
$G$ are connected subgraphs of $G$ augmented with two edges at the end, i.e., path states $(x, i, i+1, i+2, \ldots, j, y)$ distributed over $(x, \tau(i), \tau(i+1), \tau(i+2), \ldots, \tau(j), y)$. (See Theorem~\ref{thm:path-one} for 
the \bluee{rationale}). 
We use fusion operations sufficient to build the above states iteratively, starting from
the basic link-EP states. 
This set of intermediate states and fusion operations over them--yields the hypergraph used
to develop the linear program for the \gsg problem. 
We start by developing the notation used to define the intermediate states above.

\begin{figure}[h]
    \centering
    \includegraphics[width=0.48\textwidth]{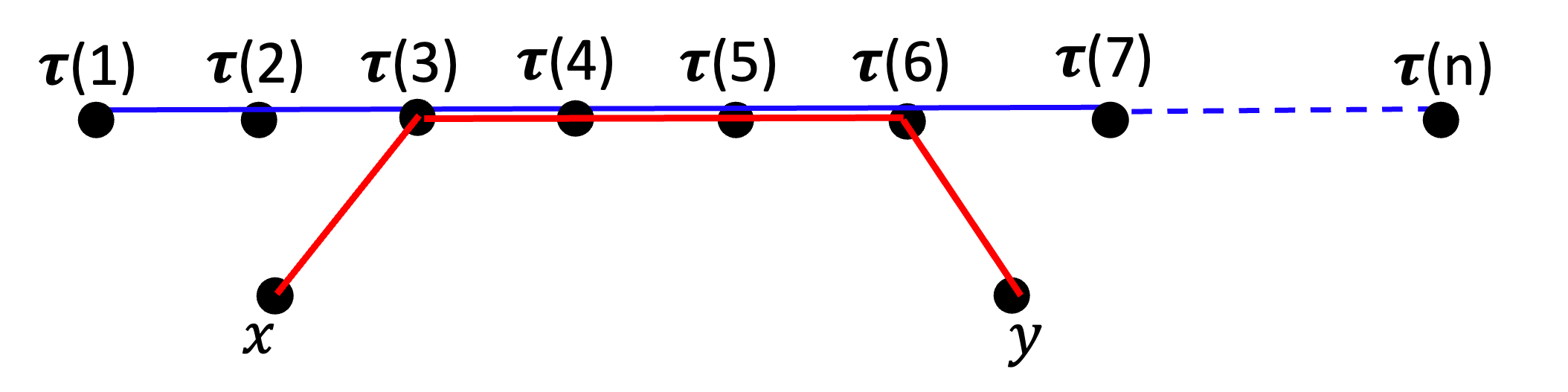}
    \caption{Notation $\langle x, 3 \cdotp \cdotp 6, y \rangle$ represents a 
    path state $(x, 3, 4, 5, 6, y)$ distributed (shown in red) 
    over the network nodes $(x, \tau(3), \tau(4), \tau(5), \tau(6), y)$. The target   state $G = (1, 2, \ldots, n)$ with distribution function $\tau()$ is in blue.}
    \label{fig:path-notation}
\end{figure}

\para{Notation $\langle x, \idj, y \rangle$.}
Recall that the target graph state $G$ is a path state $(1, 2, 3 \ldots, n)$ 
with the distribution function $\tau()$.
We use the notation $\langle x, \idj, y \rangle$, where
$1 \leq i \leq j \leq n$ and $x, y$ are vertices in the QN, 
to represent the path state $(x, i, i+1, i+2, \ldots, j, y)$ distributed
over the network nodes $(x, \tau(i), \tau(i+1), \tau(i+2), \ldots, \tau(j), y)$. 
See Fig.~\ref{fig:path-notation}.
The above notation is versatile: $i$ may be equal to $j$, signifying a path graph state $(x, i, y)$; or, the middle parameter $\idj$ may be null (\nn), signifying an edge graph state $(x,y)$; or, $x$ and/or $y$ may be null. 
To avoid duplicates, we enforce that if $i=j$ or $\idj$ is \nn, 
then $\tau'(x) \leq \tau'(y)$ \nblue{for a distribution mapping $\tau'$.}
\eat{Single qubits are valid graph states; however they may never be generated.}

\para{Intermediate States.}
As mentioned above, for a given target path graph state $(1, 2, \ldots, n)$, we choose the
following set of (distributed) {\it intermediate states}: 
$\langle x, \idj, y \rangle$ with $i,j \in [1,n]$, 
and $x$ and $y$ being any network nodes. Thus, the total number of intermediate states is approximately $n^2|V(Q)|^2$.

\begin{figure*}[ht]
    \centering
    \includegraphics[width=0.95\textwidth]{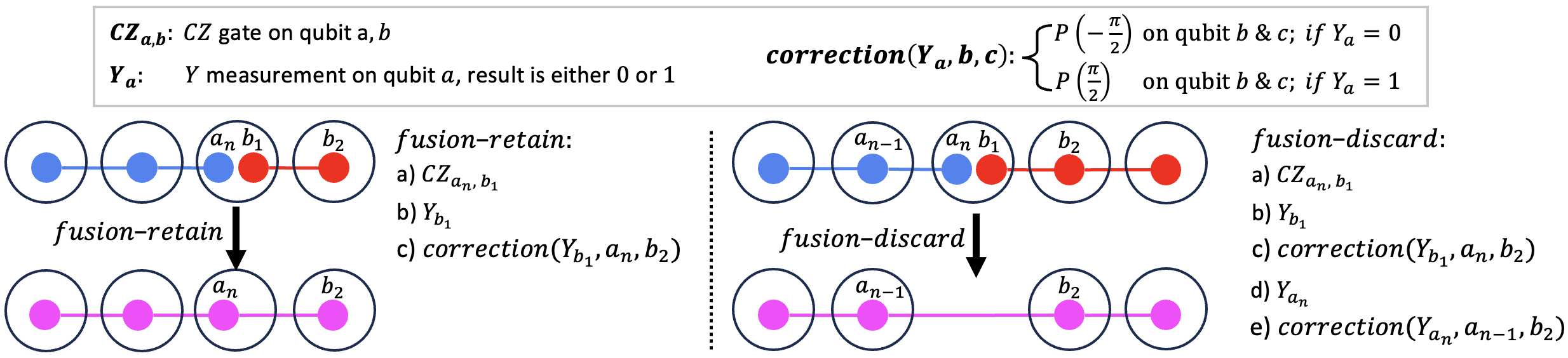}
    \vspace{0.1in}
    \caption{Fusion-retain and Fusion-discard operations. They consist of local operations: a $CZ$ gate, $Y$ measurement(s), and phase-shift operations $P()$.}
    \vspace{-0.2in}
    \label{fig:fusion-retain}
\end{figure*}

\para{Fusion-Retain and Fusion-Discard Operations.}
We use fusion operations, viz., {\it fusion-discard} and {\it fusion-retain}, to manipulate
path graph states. The fusion-discard operation merges path graph states $(a_1, a_2, \ldots, a_n)$ and $(b_1, b_2, \ldots, b_m)$ to create $(a_1, a_2, \ldots, a_{n-1}, b_2, b_3, \ldots, b_m)$,
if $a_n$ and $b_1$ are mapped to (i.e., reside at) the same network node. 
The fusion-retain operation merges path graph states $(a_1, a_2, \ldots, a_n)$ and $(b_1, b_2, \ldots, b_m)$ to create $(a_1, a_2, \ldots, a_{n}, b_2, b_3, \ldots, b_m)$,
if $a_n$ and $b_1$ are mapped to the same network node. See Fig.~\ref{fig:fusion-retain}, which also shows the local operations used in the fusion-retain and fusion-discard operations.

\para{Hypergraph.}
\bluee{We now construct a hypergraph with the above intermediate states as vertices, with the fusion operations over these vertices yielding the hyperedges, as formally described below. Such a hypergraph embeds all level-based structures that represent a generation of the given target graph state.}

\softpara{Hypergraph Vertices.}
The hypergraph consists of the following 
vertices.
\begin{enumerate}
  \item Two distinguished vertices \emph{start} and \emph{term}.
  \item $Avail(x,\idj,y)$
  for each intermediate state $\langle x, \idj, y \rangle$.
  \item $Prod^r(x,\idj,y)$, and $Prod^d(x,\idj,y)$ vertices for the $Avail(x,\idj,y)$ network nodes, as described below.\footnote{\blue{$Prod()$ signifies the graph states {\em produced} from fusion operations; the generation rate of these
  states is further reduced, to account for the fusion-success rate, before being made available in the form of $Avail()$ for further fusions.}}
\end{enumerate}

\begin{figure}[h]
    \centering
    \includegraphics[width=0.48\linewidth]{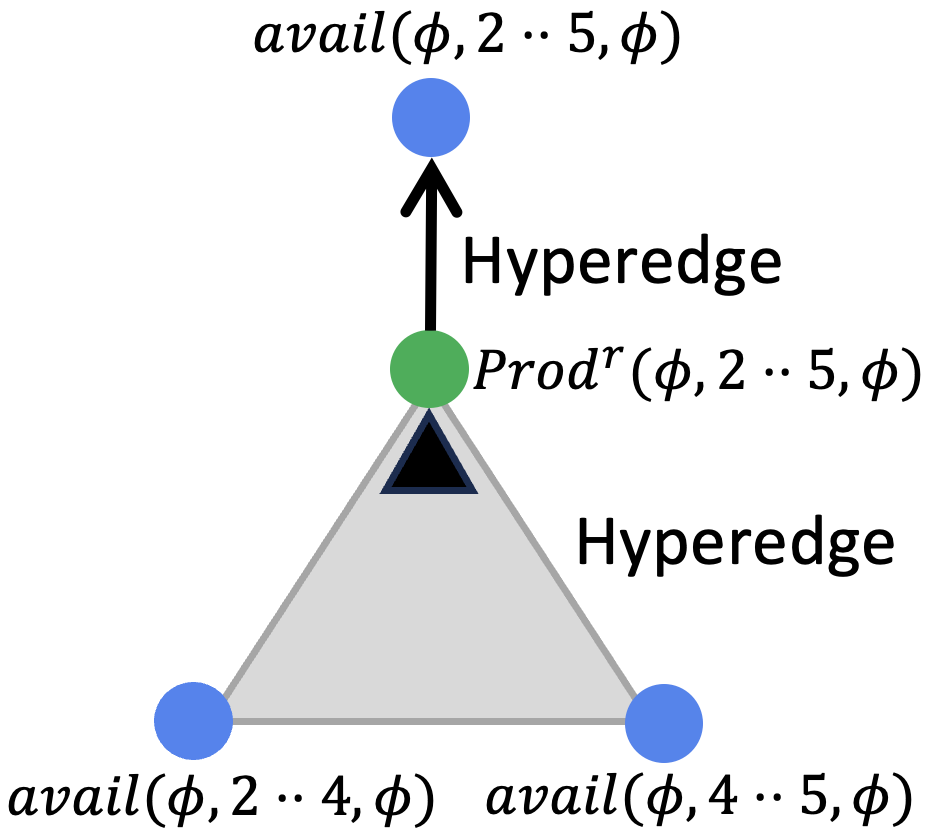}
    \caption{Two hyperedges created to represent a fusion-retain operation.}
    \label{fig:hyper-edge}
\end{figure}

\softpara{Hypergraph Edges.} 
The hyperedges {\it should} intuitively be of the type $(\{Avail(s_1),$ $Avail(s_2)\}$, $Avail(s_3))$, signifying the fusion of
states $s_1$ and $s_2$ to generate $s_3$. 
However, to incorporate the stochasticity of the fusion
operations, we create {\it two} hyperedges: 
$(\{Avail(s_1), Avail(s_2)\}, Prod^f(s_3))$ and $(Prod^f(s_3),
Avail(s_3))$,\footnote{When the hyperedge's head is singleton, 
we omit the brace brackets. \bluee{Also, {\it prod} signifies production, while {\it Avail} signifies available for consumption.}}
where the first edge represents the fusion 
operation $f$ while the second edge incorporates the fusion's
probability of success \bluee{(see Eqn.~\ref{eqn:prod}). See Fig.~\ref{fig:hyper-edge}.}
In our context, the superscript $f$ over $Prod()$ 
is $d$ ($r$) for fusion-discard (fusion-retain).
Overall, we have the following set of hyperedges. 

\begin{figure}[h]
    \includegraphics[width=0.5\textwidth]{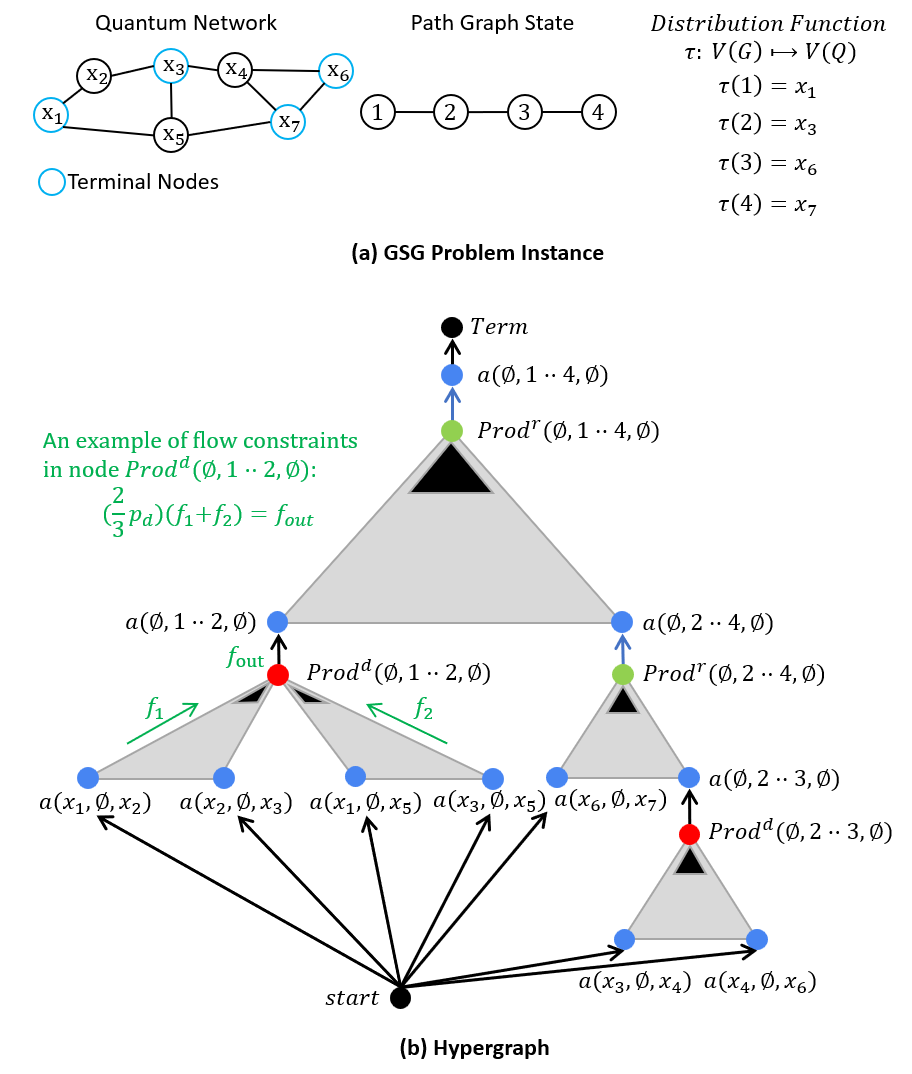}
    \caption{(a) A \gsg problem instance for a path graph state, and (b) A corresponding hypergraph (for sake of clarity, we have shown only a small subset of relevant vertices and hyperedges); here, $a() = Avail()$.}
    \label{fig:path-lp}
\end{figure}

\begin{enumerate}
  \item
  Hyperedges $(start, avail(x, \nn, y))$ for all network links $(x,y)$, representing 
 \bluee{generation of link states directly from the network nodes.}
  
  \item{[$fusion\_discard$] hyperedges}.
  We create hyperedges to represent a generation of intermediate states from other states via the fusion-discard operation described above. E.g., by fusing 
  states $\langle x, \idj, z \rangle$ and $\langle z, (j+1)\cdotp \cdotp k, y \rangle$, we get
   $\langle x, \idk, y \rangle$. Thus, we create the hyperedges: 
  \begin{itemize}
  \begin{small}
  \item  $(\{Avail(x,\idj,z), Avail(z, {(j+1)}\cdotp \cdotp k,y)\}, Prod^d(x,\idk,y))$
  \item   $(Prod^d(x,\idk,y), Avail(x,\idk,y))$
  \end{small}
  \end{itemize}
 We must also create pairs of hyperedges corresponding to intermediate states with null (\nn) parameter values. E.g., $\langle x, \nn, z \rangle$ and $\langle z, \nn, y \rangle$ can be fused to get $\langle x, \nn, y \rangle$. We omit stating these cases here for clarity of presentation. 
 
\item{[$fusion\_retain$] hyperedges.} Similarly, we create the following hyperedges due to fusion-retain operations. See Fig.~\ref{fig:hyper-edge}.
\begin{itemize}
\begin{small}
         \item $(\{Avail(x,\idj,\nn), Avail(\nn,\jdk,y)\}, Prod^r(x,\idk,y))$
         \item $(Prod^r(x,\idk,y, Avail(x,\idk,y))$
         \item $(\{Avail(x,\idi,\nn), Avail(y,\idi,\nn)\}, Prod^r(x,\idi,y))$
         \item $(Prod^r(x,\idi,y, Avail(x,\idi,y))$ 
\end{small}
\end{itemize}
\item
\bluee{Hyperedge $(Avail(\nn, 1\cdotp\cdotp n, \nn), term)$, signifying generation of the target graph state.}
\end{enumerate}
See Fig.~\ref{fig:path-lp} for an example hypergraph.

\para{LP Variables.} For each hyperedge $e = ((u,v), w)$, 
we create an LP variable $z_e \in R^+$ which 
represents the rate of the fusion operation (and thus, its operands and result). 
This implicitly enforces the 
\bluee{(desirable)} 
condition
that the generation 
rates of the states/vertices $u$ and $v$ {\it used for any edge} $e$ are equal.

\para{LP Constraints and Optimization Objective.}
\begin{itemize}
    \item \textbf{Capacity Constraints:} Each network node $x$ has an atom-photon generation capacity constraint.
     $$  \hspace*{-0.2in} 1/t_g \geq \sum_{(x,y) \in E(Q)}  z_{(\{start\}, Avail(x,y))} / (p_g^2 p_e^2 p_{ob}) \ \ \forall x \in V(Q)  $$

    \item \textbf{Flow Constraints:} Flow constraints vary with vertex types. Let 
    $out(v)$ and $in(v)$ denote the set of outgoing and incoming hyperedges from a vertex $v$ in the hypergraph. Formally, $out(v)$ is  $\{e\in E(H) \ | \ v \in t(e) \}$, and $in(v)$ is $\{e\in E(H) \ | \ v = h(e) \}$.

    \begin{itemize}
        \item For each vertex $v$ s.t. $v=Avail()$:
              $$ \sum_{e \in in(v)}  z_e  = \sum_{e' \in out(v)} z_{e'} $$
             
        \item For each vertex $v$ s.t. $v=Prod^r(\cdot)$:
        \begin{equation}
            \sum_{e \in in(v)} (\frac{2}{3} p_r)z_e = \sum_{e' \in out(v)} z_{e'} 
            \label{eqn:prod}
        \end{equation}    
              Here, $p_r$ is the probability of success of the fusion-retain operation. 
        \item For each vertex $v$ s.t. $v=Prod^d(\cdot)$:
              $$ \sum_{e \in in(v)}  (\frac{2}{3}p_d) z_e  = \sum_{e' \in out(v)} z_{e'} $$
              Here, $p_d$ is the probability of success of the fusion-\nblue{discard} operation.
              
    \end{itemize}

    \item \textbf{Objective}: We maximize the sum of the generation rates of the hyperedges incoming into the {\it term} vertex. 
    $$ \max\ \sum_{e \in in(term)} z_e$$.
\end{itemize}

\vspace*{-0.1in}
\para{Optimality Result.}
\blue{LPs can be solved optimally in polynomial time using interior-point methods or simplex-based approaches. In our evaluations, we use the Gurobi Solver to solve the LPs arising from our problem formulation.}
We can show the below optimality result. 
\cbb

\begin{thm-prf}
The above LP-based algorithm 
returns an optimal level-based structure for the special-case
of the \gsg problem wherein the target graph state $G$ is a path graph, and the output level-based structure $L$ is such that: 
(a) The leaves of $L$ corresponding to all link-EPs;
(b) The interior nodes of $L$ corresponding to a (intermediate) distributed graph state 
$H$ (with a mapping $\tau$) with the following restrictions: 
(i) The mapping $\tau$ is onto, i.e., each qubit in $H$ maps to a unique network node,
(ii) $H$ is a connected subgraph (not necessarily induced) of $G$ with additional ''leaf'' edges (i.e., edges with one of the vertices having a degree of one);
(c) The fusion operations used in $L$ are fusion-discard and fusion-retain, with the restriction that the fusion-retain operation is used only at terminal nodes over qubits with degree one in their graph states.
\label{thm:path-one}
\end{thm-prf}
\begin{prf}
First, we note that an LP formulation can be optimally solved in polynomial time.
Thus, we only need to prove that the {\em hypergraph} used to formulate the above-described LP formulation considers all the intermediate graph states and
fusion operations among them, under the given restrictions. This, in turn, would
imply that any level-based structure $L$ that satisfies the given restrictions is
captured by the LP formulation --- thus, proving the optimality of the LP-based algorithm.

We can prove by induction (on the number of fusion operations) 
that any intermediate state formed by the application of the given fusion operations 
can be represented by our path notation $\langle x, \idj, y \rangle$ 
and thus is a vertex in the LP's hypergraph. The base case is trivially
true, since the notation can represent the link-EPs. Now, lets
consider the given fusion operations applied to two states of the form
$\langle x_1, \idj, y_1 \rangle$ and $\langle x_2, r \ldots s, y_2 \rangle$
with mappings $\tau_1$ and $\tau_2$ respectively. 

(a) \underline{Fusion-Retain Operation.} Let's first consider the 
fusion-retain operation. Let $\tau_1(j) = \tau_2(r)$, so that the
fusion-retain can be applied to these two qubits at the node $\tau(j)$. 
This also means that $j$ must be equal to $r$. As per the given restriction, $y_1$ and $x_2$ must be null. Then, the fusion-retain
operation leads to the state $\langle x_1, i \ldots s, y_2 \rangle$ with an appropriately
defined mapping $\tau_3$. Note that the fusion-retain operation can only be applied to the qubits at terminals and that if $\tau_1(i) = \tau_2(r)$, then application of fusion-retain at $\tau_1(i)$ would need
to the violation of the requirement that intermediate states should only have a single
qubit at each node.

(b) \underline{Fusion-Discard Operation.} 
Now, let's consider the fusion-discard operation. First, we note that the
fusion-discard operation can't occur at the terminal nodes since the fusion-discard at $\tau_1(j) = \tau_2(r)$ (with $j = r$, also) would lead to a state with edge $(j-1, r+1) \notin H$; similarly, fusion-discard at a node $\tau_1(k)$, where $i < k < j$ would lead to edges not in $H$. 
Now, fusion-discard at $y_1 = x_2$ leads to the state: 
$\langle x_1, i \ldots s, y_2 \rangle$ with an appropriate mapping; note that, 
in this case, $j$ must be equal to $r-1$. This holds for all possible cases, viz., when $i=j$ and/or $r=s$, or $x_1 = \nn$ or $y_2 = \nn$.

Finally, it is easy to see that the hyperedges in the hypergraph associated with the LP capture all the fusion operations allowed without violating the given conditions. 
\end{prf}

The theorem and its proof can 
be generalized to allow for more general fusion operations; 
in addition, the restriction on fusion-retain operations in the above theorem can also be relaxed but requires tedious analysis. We omit these details for 
clarity of presentation. 

We also note that the performance of the LP-based solution depends 
greatly on the network topology and other parameters; e.g., if the
network topology itself is a path graph, then even a simpler dynamic-programming solution (similar to the one in~\cite{swapping-tqe-22} for EP generation) would be optimal. 
\cbl

\subsection{\textbf{Computationally-Efficient LP Formulations}}
\label{sec:path-apx}

Even though the above LP formulation is polynomial-time and returns an optimal \gsg solution, 
it can be computationally prohibitive for even moderate-size networks. E.g., for a network of 100 nodes and a path graph state of 10 terminals, the number of intermediate states is about a million, and the LP consists of 100s of millions 
of variables (from that many hyperedges). 
Such LP formulations can be computationally infeasible. 
Thus, we develop the below LP formulations that sacrifice optimality for computational efficiency. In each of the below schemes, the hypergraph is an {\it induced} sub-hypergraph of the hypergraph from \S\ref{sec:opt-path}. Thus, defining the set of intermediate states (and, thus, the hypergraph vertices) for a scheme is sufficient for its full description.

\para{Distance-Based LPs.}
In this class of LP formulations, we only consider the intermediate states $\langle x, \idj, y \rangle$, where the node $\tau'(x)$ is within a certain
distance (physical \blue{and/or hop-count})\footnote{\blue{Technically, we should use a combination of physical distance {\em and} hop-count, as both metrics have an impact on the generation rate. In particular, the hop count directly affects the generation gate due to the number of swappings involved, and a longer physical distance {\em can} independently entails longer physical distances over individual links in the entanglement path. However, in our evaluations, we only used physical distance since our network instances were spread over a constant area.}} from the 
terminal $\tau'(i)=\tau(i)$. The intuition is that intermediate states 
$\langle x, \idj, y \rangle$ where $\tau'(x)$ is very far away from $\tau(i)$ 
is unlikely to be helpful in an efficient 
generation of $G$. More formally, we impose the condition:
$ d(\tau'(x), \tau(i)) \leq c \cdot \max (d(\tau(i-1), \tau(i)), d(\tau(i), \tau(i+1))),$
where $d()$ is an appropriate distance function and $c > 0.5$.

\para{Left-Sided and Right-Sided LPs.}
In this scheme, we only consider intermediate states of the type $\langle x, \idj, \nn \rangle$. Similarly, we can consider a scheme that only considers states $\langle \nn, \idj, y \rangle$. 
We refer to these schemes as {\tt Left-Sided LP} and {\tt Right-Sided LP} respectively. 

\begin{figure}[h]
    \centering
    \includegraphics[width=0.48\textwidth]{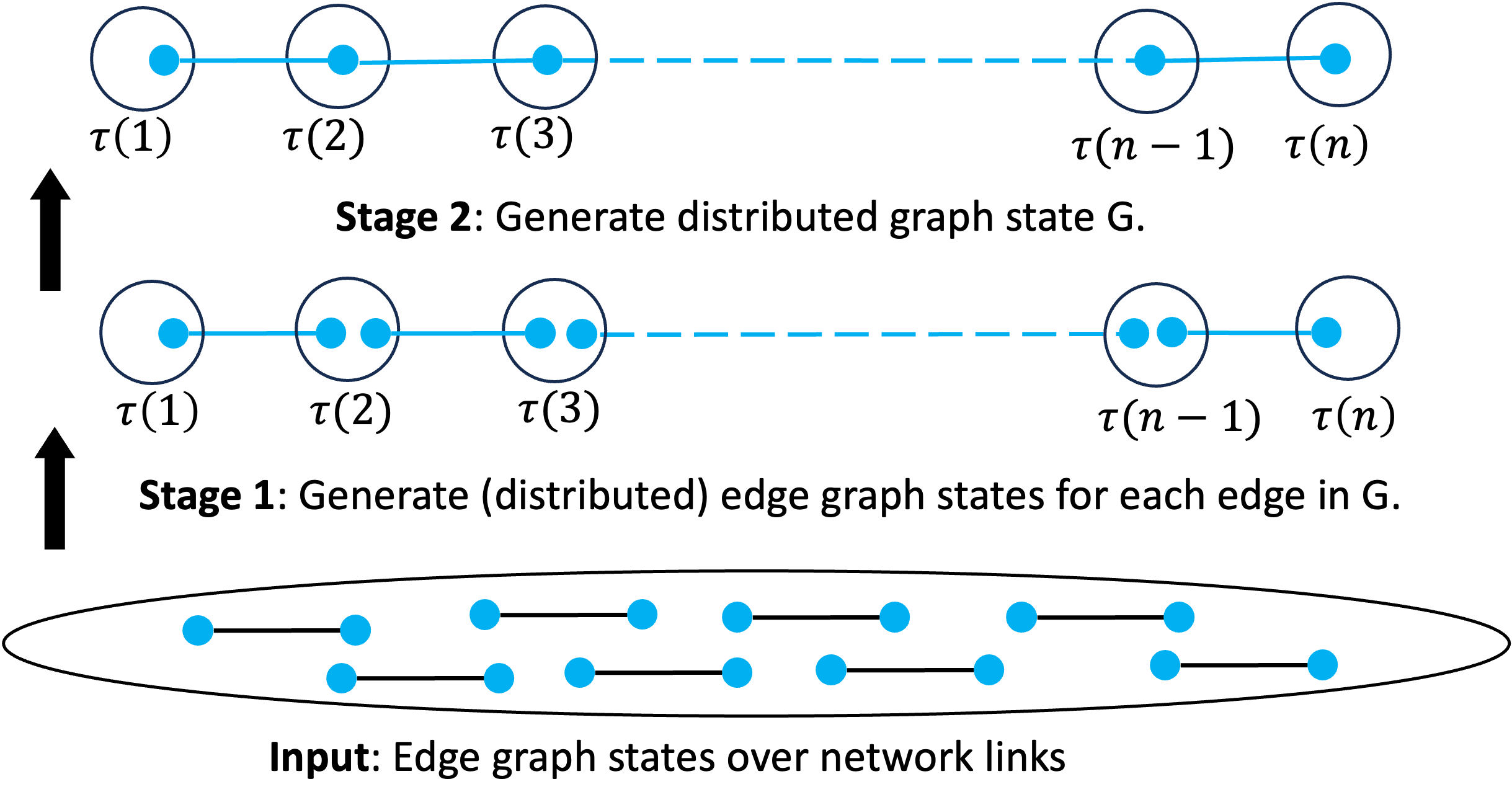}
    \caption{Two-Stage Generation Scheme for a Path Graph State $(1, 2, \ldots, n)$ with a distribution mapping $\tau$.}
    \label{fig:two-stage}
\end{figure}

\para{Two-Stage LP.}
In the {\tt Two-Stage LP} (see \S\ref{sec:high-level}), we 
generate the target path graph state in two stages. In the first stage, we 
create the single-edge graph states $\langle \nn, i \cdotp \cdotp {(i+1)}, \nn \rangle$ 
for all $i \in [1, n-1]$---using the link states and other edge states created
in this stage. 
Then, in the second stage, we create (intermediate) states of the type: 
$\langle \nn, \idj, \nn \rangle$, eventually yielding the target graph 
state $\langle \nn, 1 \cdotp \cdotp n, \nn \rangle$---using only the 
first-stage edge graph states and second-stage states (and thus, not 
involving any of the non-terminal nodes). 
\bluee{Note that the states considered in the second state are all 
the connected subgraphs of the target path state, which are $O(n^2)$.}
See Fig.~\ref{fig:two-stage}.
Another way to look at the above {\tt Two-Stage} scheme is as follows: 
Consider the induced subgraph of the hypergraph from~\S\ref{sec:opt-path} 
over the vertices of the 
type $\langle \nn, \idj, \nn \rangle$ or $\langle x, \nn, y \rangle$.

\para{Performance Guarantees.} We can show the following, \blue{similar to the proof of Theorem~\ref{thm:path-one}.} 
\begin{theorem}
The above {\tt Two-Stage} scheme returns an optimal solution for the special case
of the \gsg solution mentioned in Theorem~\ref{thm:path-one} with
\bluee{the additional requirement that the level-based structure $L$ has a ``barrier''  level (i.e., no state at higher level depending on states at lower levels) consisting only of single-edge states corresponding to the edges in $G$.}
\label{thm:path-two}
\end{theorem}

\section{\bf Generating Tree Graph States}
\label{sec:general_fusion}

We now design efficient generation schemes to generate 
tree graph states.
Unlike for path graphs, the number
of connected induced subgraphs of a tree is exponential 
in the number of vertices. 
Thus, considering all connected
induced subgraphs (e.g., as in \S\ref{sec:opt-path} for paths) is not
feasible. In this section, we design two schemes 
based on a combination
of strategies to reduce the number of intermediate states considered.

\subsection{\textbf{Two-Stage Generation Scheme}}
\label{sec:tree-two}

Consider a \gsg problem instance, wherein the target graph state $G$ is a tree $T$.
Recall that the target distribution of $G$ over $Q$ is represented by $\tau: V(G) 
\mapsto V(Q)$.


\para{Basic Idea.}
As described in previous sections, a Two-Stage approach 
consists of two stages. In the first stage, we generate
single-edge states corresponding to edges in the target state, and then, 
in the second stage, we iteratively generate appropriate types of 
intermediate states and, eventually, the target state.
Generally, the natural set of intermediate 
states to consider in the second stage is the set of 
all connected subgraphs of the target state (as in \S\ref{sec:path-apx} for paths).
However, for a tree state, that
is exponential. Thus, we consider a carefully chosen set of 
specialized connected subgraphs such that they are polynomial 
in number, can be computed from link states via other states 
from this set (in other words, the set of states yields a connected 
hypergraph), and is ``rich'' enough to facilitate an efficient LP 
solution. We start with a notation 
that defines these specialized subgraphs of trees.

\begin{figure}[h]
    \centering
    \includegraphics[width=0.35\textwidth]{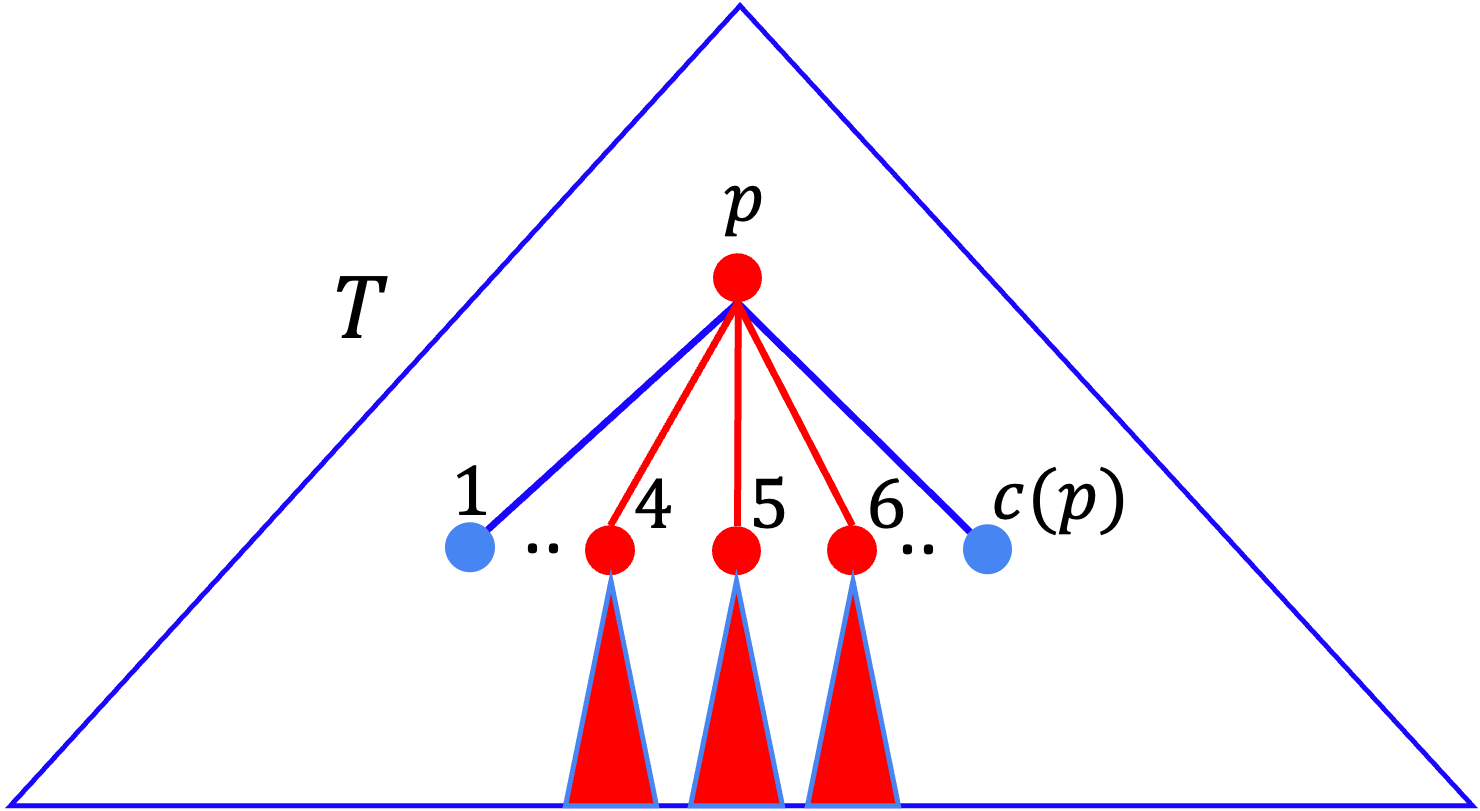}
    \vspace{0.1in}
    \caption{Notation $\texttt{Tree}(p,\idj)$ (here, $i=4, j=6$) denotes a distributed graph state $T'$ (shown in red) that is an induced subgraph of $T$ including the interior node $p$, the $i^{th}$ to $j^{th}$ children of $p$ and all their descendants. 
    The graph state $T'$ has the same distribution function $\tau()$ as $T$.} 
    \label{fig:tree-state}
\end{figure}

\para{Notation \texttt{Tree}$(p,\idj)$.} 
Consider a \gsg problem instance: a quantum network (QN) $Q$ and a tree graph state 
$T$ along with its target distribution $\tau: V(T) \mapsto V(Q)$. 
Without loss of generality, we {\it number} the children of each non-leaf node 
$x$ in $T$ from $1$ to $c(x)$, where $c(x)$ is the number of children of $x$ in $T$.
Based on such a numbering, the notation \texttt{Tree}$(p, \idj)$, where $p \in V(T)$ 
and $1 \leq i \leq j \leq c(p)$, denotes a distributed tree state $T'$ that is an
induced subgraph $T'$ of $T$ containing 
the following vertices: 
(i) node $p$ as $T'$ root, 
(ii) $p$'s $i^{th}$ to $j^{th}$ children along with {\it all} their
descendants in $T$. In addition, $T'$ also uses the same distribution 
function $\tau$ over its vertices. See Fig.~\ref{fig:tree-state}.

\softpara{Remark.} 
\blue{Note that the above tree notation is specifically designed to represent all and only the intermediate states 
that can arise during the two-stage generation scheme for a 
given tree graph state. 
(Later, in \S\ref{sec:tree-one-stage},
we extend/modify our notation 
to represent intermediate states that can arise in the {\em one-stage}
generation scheme.) 
Moreover, the only fusion operations permitted in our schemes are the ones explicitly defined here. Thus, other fusion operations, e.g., fusing one tree's leaf with another tree's root, are not allowed.}

\begin{figure}[h]
    \centering
    \includegraphics[width=0.45\textwidth]{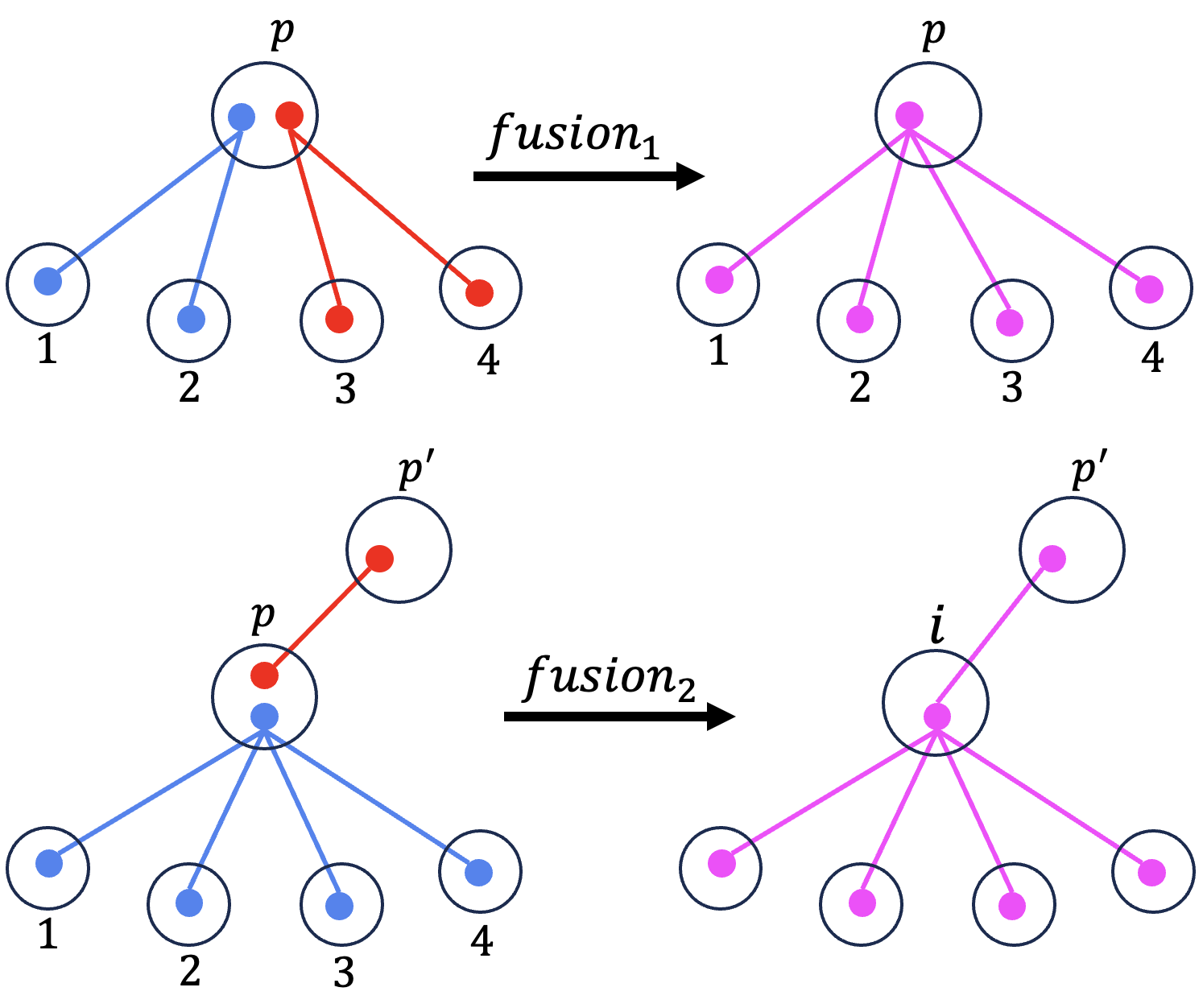}
    \caption{$Fusion_1$ and $Fusion_2$ operations. Here, the numbers 1 to 4 represent the children numbers of the parent node/vertex $p$.}
    \label{fig:tree-fusions}
\end{figure}

\para{Overall Two-Stage Scheme.}
Our Two-Stage generation scheme for the tree graph states consists of two stages.
\begin{itemize}
\item 
First, from the link states, we generate single-edge graph states corresponding
to each edge in $T$. 
The intermediate states considered in this stage are 
single-edge distributed graph states for all pairs of network nodes.

\item 
Then, using the above single-edge states (output of the first stage) and other states 
generated in this second stage, we iteratively generate intermediate states (which include the
target state) of the kind $\texttt{Tree}(p,\idj)$ where $p \in V(T)$ and $1 \leq i \leq j 
\leq c(p)$.
\end{itemize}
For the first stage, we need to use only the {\it fusion-discard} operation (see \S\ref{sec:opt-path}), while for the second stage, we use the following fusion operations (see Fig.~\ref{fig:tree-fusions}).

\begin{itemize}
\item 
$Fusion_1$: Fuse \texttt{Tree}$(p,\idj)$ and \texttt{Tree}$(p, {(j+1)} \cdotp \cdotp k)$ to generate \texttt{Tree}$(p,\idk)$.

\item 
$Fusion_2$: Fuse \texttt{Tree}$(p, 1 \cdotp \cdotp c(p))$ and $\langle p, p' \rangle$ (with $p$ and $p' \in T$ mapped to $\tau(p)$ and $\tau(p')$ respectively) to generate \texttt{Tree}$(p', \idi)$; here, $p$ is the $i^{th}$ child of $p'$ in $T$.
\end{itemize}

\para{Hypergraph and LP.}
We now construct a hypergraph with $Avail()$ vertices for all link states and 
intermediate states (from both stages), and hyperedges to represent the fusion operations 
over the $Avail()$ vertices as described above.
As in the previous section, we add $Prod^*$ vertices for each fused $Avail()$ node 
based on the fusion operation used. 
We formulate the constraints
and the objective function in the LP, as in the previous section. 
We state the performance guarantee in Theorem~\ref{thm:tree},
in the following subsection.

\subsection{\textbf{One-Stage Generation Scheme}}
\label{sec:tree-one-stage}

To improve the above Two-Stage LP formulation, 
we add vertices (and corresponding hyperedges) 
to the hypergraph of the previous subsection 
to ``bridge the separation'' between the two stages. 
In particular, 
we expand the previous set of 
intermediate 
states by allowing an arbitrary network node to have 
\texttt{Tree}$(p,\idj)$ as its subtree. 
The new set of intermediate states is still polynomial
in input size, but ``connects'' the first-stage 
intermediate states to other stages in the hypergraph of
Two-Stage approach and thus enabling a richer 
set of hyperpaths and level-based 
structures in the LP.

\para{Notation \texttt{Tree}$(x, p,\idj)$}. 
This denotes a distributed tree graph state $T'$ that includes a 
vertex $x$ (corresponding to an arbitrary network node) as
the root, with its sole child as the tree graph state 
\texttt{Tree}$(p,\idj)$. $T'$ distribution function $\tau'$ is 
same as $\tau$ for $p$ and its descendants, and for $x$, 
$\tau'(x) = x$.

\begin{figure}[t]
    \centering
    \includegraphics[width=0.45\textwidth]{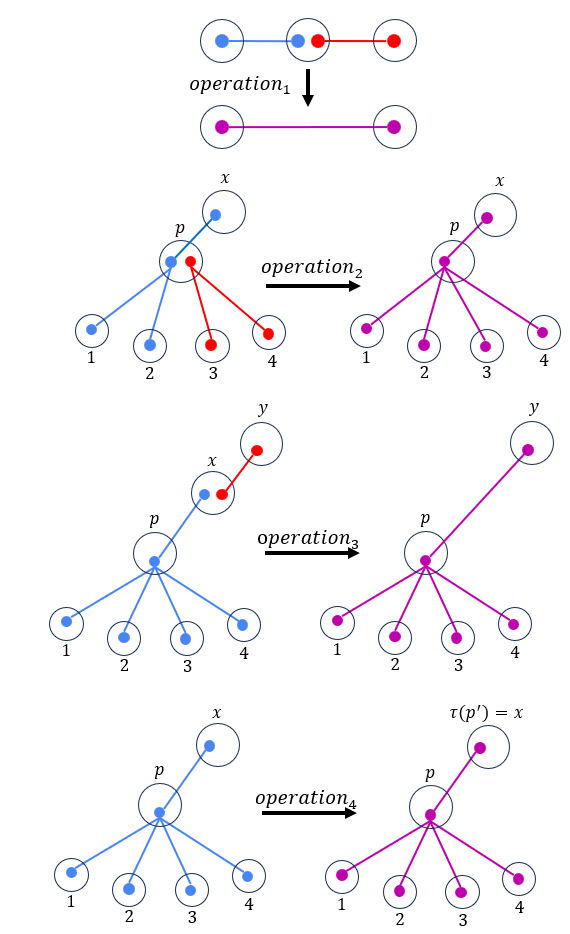}
    \caption{\blue{The four operations for the One-Stage generation scheme for tree graph states. Here, the numbers 1 to 4 represent the children numbers of the parent node/vertex $p$.}}
    \label{fig:tree-fusions_2}
\end{figure}

\para{Intermediate States, Fusion Operations, Hypergraph.}
We select the set of intermediate states as all states of the type 
(i)  \texttt{Tree}$(x, p,\idj)$ with 
$\tau'(x) \in V(Q), p \in V(T)$ and $1 
\leq i \leq j \leq c(p)$, and (ii) 
Single-edge graph states, for
every pair of network nodes; we denote these states by $\langle y, z \rangle$ 
where $y,z$ are network nodes.
We use essentially the similar fusion operation as for the two-stage scheme, except that we 
also add fusion operations to allow the extension $x$ of $p$ in \texttt{Tree}$(x, p,\idj)$  
to extend so that $x$ is mapped to $\tau(p')$, at which point, 
the distributed state transforms to 
\texttt{Tree}$(\red{\nn}, p', \idi)$. More formally, we allow the following fusion operations \blue{(see Fig.~\ref{fig:tree-fusions_2})}: 
\begin{enumerate}
\item 
Fusion-discard operation over edge graph states, i.e., fuse states $\langle x, y \rangle$ and $\langle y, z \rangle$ to form $\langle x, z \rangle$.

\item
Fuse states \texttt{Tree}$(x, p,\idj)$ and \texttt{Tree}$(\nn, p, {(j+1)} \cdotp \cdotp k)$ to generate \texttt{Tree}$(x, p,\idk)$, and similarly, 
\texttt{Tree}$(\nn, p,\idj)$ and \texttt{Tree}$(x, p, {(j+1)} \cdotp \cdotp k)$ to generate \texttt{Tree}$(x, p,\idk).$ These are similar to $Fusion_1$ in the Two-Stage scheme, but with the $(x,p)$ extension.

\item 
Fuse states \texttt{Tree}$(x, p,\idj)$ and $\langle y, x \rangle$ to generate \texttt{Tree}$(y, p,\idj)$. This is essentially the fusion-discard operation to extend the extension $(x,p)$.

\item 
Transform (without any fusion) 
\texttt{Tree}$(x, p, 1 \cdotp \cdotp c(p))$ to  
\texttt{Tree}$(\nn, p', \idi)$ where $p$ is the $i^{th}$ child of $p'$ in $T$ and 
$\tau(p') = x$. 
\end{enumerate}

Based on the above intermediate states and the fusion operations, we construct a hypergraph $H_1$ and formulate an LP as before. 
It is easy to see the following: (i) The hypergraph for the Two-Stage scheme (\S\ref{sec:tree-two}) is an induced subgraph of the above-constructed hypergraph $H_1$. (ii) There are level-based structures in $H_1$ 
that do not use any edge graph states corresponding to $T$'s edges---which means that, in this One-Stage scheme, the target graph state $T$ can be potentially generated without going through any edge graph states corresponding to $T$'s edges. (iii) The total number of intermediate states in the above One-Stage scheme is polynomial in the size of the network and the target graph state. \blue{The following theorem holds for both the schemes for tree graph states, largely from 
the fact that an LP formulation can be solved optimality.}

\begin{theorem}
The Two-State and One-Stage generation schemes above return an optimal solution for the special cases of the \gsg problem wherein (a) the target state is a tree graph, (b) the output level-based structure $L$ is such that the vertices and fusion operations used in $L$ are restricted to the intermediate states and fusion operations discussed above, in the respective schemes.
\label{thm:tree}
\end{theorem}

\section{\bf Other Graph States; Multiple Graphs; Fidelity}
\label{sec:gen}

\para{Generating Other Classes of Graph States.} Our LP-based technique for optimized generation of graph states is very versatile; it can be tailored to generate other classes of states. 

\begin{figure}
    \centering
    \includegraphics[width=0.45\textwidth]{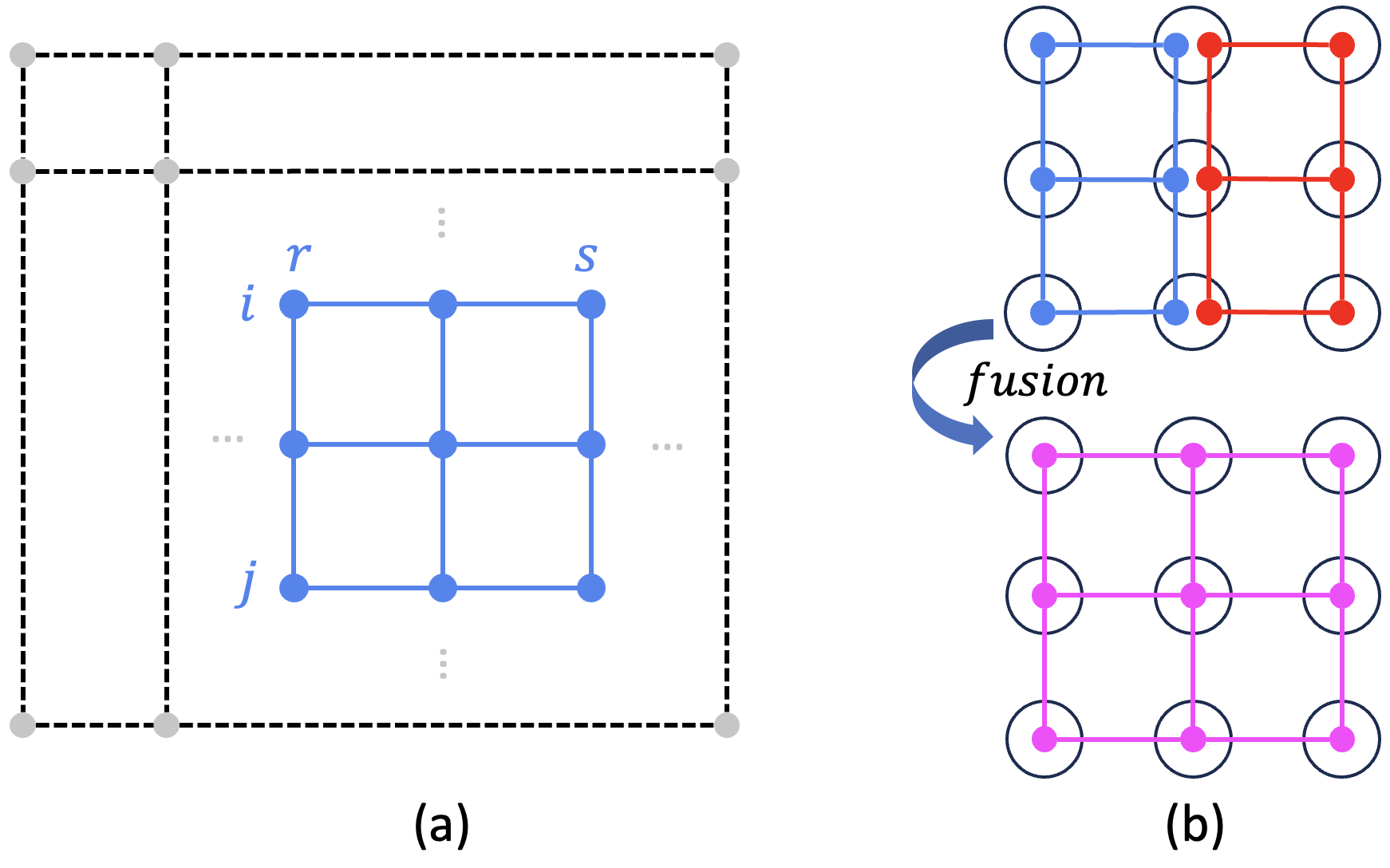}
    \caption{Grid graph states. (a) Intermediate state. (b) Fusion-Column Operation.}
    \label{fig:grid-state}
\end{figure}

\softpara{Grid Graph States.} 
A $(m_x, m_y)$-grid graph state $G$ with $m_x, m_y \geq 1$ has a 2D structure consisting of $m_x$ columns and $m_y$ rows.
For such states, it is natural to consider intermediate states of the type 
$\langle \idj, \rds \rangle$ consisting of $i$ to $j$ rows and $k$ to 
$l$ columns of $G$; the number of such states is polynomial in the size of $G$.
In addition, we must consider $\langle x, y \rangle$ single-edge graph states where $x$
and $y$ are network nodes.
To facilitate the generation of $G$ from these intermediate states, we include fusion-retain,
fusion-discard, {\it fusion-row} and {\it fusion-column} operations. 
The fusion-row operation fuses states $\langle \idj, \rds \rangle$ and $\langle \jdk, \rds \rangle$ to generate $\langle \idk, \rds \rangle$, and similarly a fusion-column operation
fuses states $\langle \idj, \rds \rangle$ and $\langle \idj, \sdt \rangle$ to create
$\langle \idj, \rdt \rangle$. Fig.~\ref{fig:grid-state} 
shows the fusion-column operation.
With the above intermediate states and fusion operations, we can construct the hypergraph 
and formulate an LP as before.

\begin{figure}[h]
    \centering
    \includegraphics[width=0.48\textwidth]{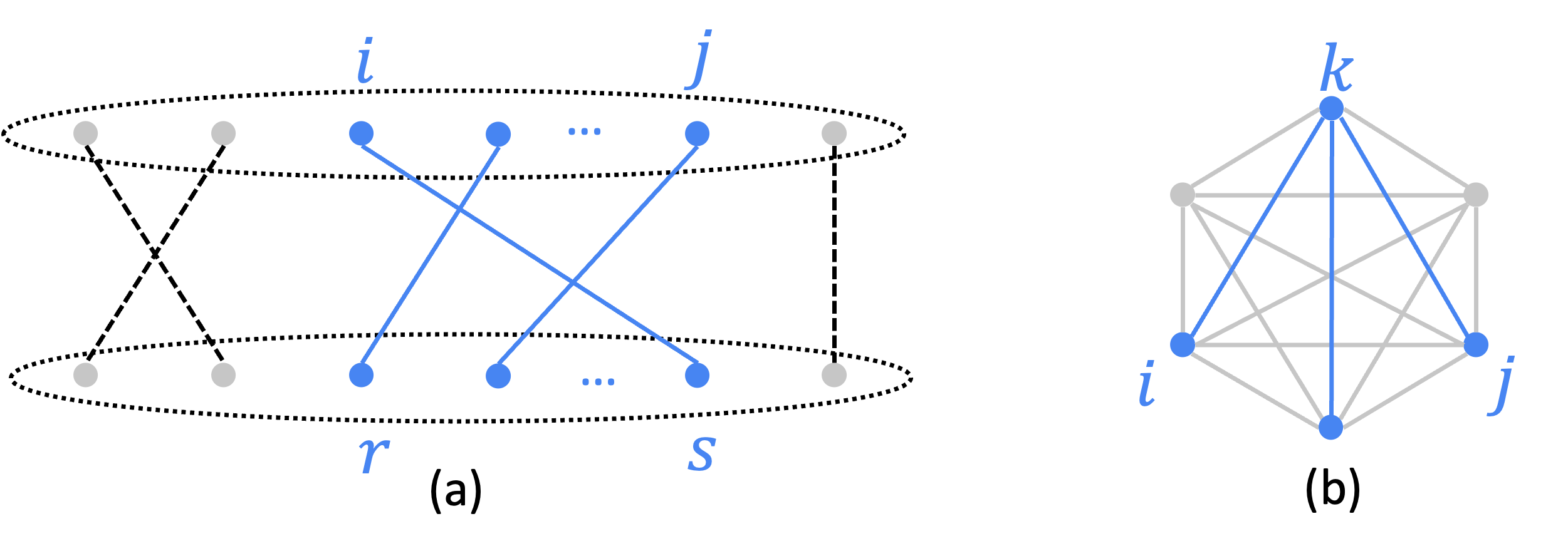}
    \caption{Intermediate states of (a) bipartite graph, (b) complete graph.}
    \label{fig:grid-state-2}
\end{figure}

\softpara{Bipartite Graph States.} A $(m_a, m_b)$-bipartite graph state $G$ 
has $m_a$ and $m_b$ vertices in the two partitions $A$ and $B$ numbered
1 to $m_a$ and 1 to $m_b$ respectively. 
To consider a polynomial number of intermediate states, we consider the intermediate states
corresponding to the induced subgraphs represented by $\langle \idj, \rds \rangle$ which includes
$i$ to $j$-numbered vertices in $A$ and $r$ to $s$-numbered vertices in $B$.
See Fig.~\ref{fig:grid-state-2}.
We include the edge graph states $\langle x, y \rangle$.
We can create appropriate fusion operations to 
generate 
$\langle \idj, \rds \rangle$ intermediate states; 
the fusion operations essentially fuse a set of star graphs. 

\softpara{Complete or Repeater Graph States.} For complete graphs $G$, we can consider the intermediate states as the star graphs $\langle k, \idj \rangle$ with vertex $k$ as its root and vertices numbered $i$ to $j$ (excluding $k$) as its children. 
Along with the single-edge graph states, the total number of intermediate states is polynomial in the size of $G$, and the only fusion operations needed are fusion-retain, fusion-discard, and a fusion
to fuse two star graphs. The above approach easily extends to repeater graphs~\cite{korean23}.

\para{Generating Multiple Graph States Concurrently.}
Our \gsg problem considers the generation of (multiple instances of) a single graph state. Our LP formulation can easily be extended to generate several different ``types'' (including different distributions of the same graph state) of graph states concurrently by essentially creating a hypergraph for each graph state, ``merging'' the hypergraphs (by taking a union of the vertices and edges, and removing duplicates), and formulating the LP formulation's objective to maximize the {\it sum} (or some linear function) of the generation rates of all the graph states. 

\para{Decoherence and Fidelity Constraints.} To incorporate fidelity and decoherence 
constraints in the \gsg problem formulation, we enforce (as in~\cite{swapping-tqe-22,ghz-qce-23})
that the structure \fusionTree 
satisfy the following constraints: (a) The number of ``leaves'' of any ``tree'' in the 
level-based structure is less than a given threshold $\fidl$; this is to 
limit fidelity degradation due to gate operations. (b) Any qubit's total 
memory storage time is less than a given \textit{decoherence threshold} 
 $\fidd$.

Theoretically, the above constraints on the loss of fidelity due to noisy fusion operations 
and the age of qubits can be added to the LP formulation as follows, in a way similar to~\cite{swapping-tqe-22} for swapping trees.
First, we observe that the fidelity degradation of a generated 
graph state due to the number of operations
can be modeled by limiting the number of its leaf descendants.
Second, as observed in~\cite{swapping-tqe-22} for the case of swapping trees,
the decoherence constraint (i.e., bounding the total age of a qubit in a 
graph state) can be incorporated by limiting the {\it depths} of the 
left-most and  right-most descendants of the children of a graph state 
in the hypergraph.
These \emph{structural} constraints can be enforced in the LP formulation
by adding the leaf count and appropriate heights as
parameters to $Prod^*$ and $Avail^*$ vertices.

\para{\nblue{Application to Quantum-Repeater based Quantum Networks.}}
\nblue{One of the fundamental services that a wide-area QN needs to provide is that of remote or long-distance entanglement generation. In general, entangled graph states have applications in various quantum information processing domains, as discussed in \S\ref{sec:intro}.
Techniques developed in this work apply to quantum network architectures with quantum 
repeaters (i.e., nodes with heralded memories and atomic-BSM or entanglement-swapping 
stations);
quantum repeaters are considered essential to enable long-distance or wide-area 
QNs~\cite{qr-1,qr-2,gyongyosi2020dynamics,gyongyosi2022advances}. 
Note that our techniques are independent of {\em how} the link-EPs are generated, which may differ across QN architectures~\cite{gyongyosi2018survey}. 
}

\section{\bf Evaluations}
\label{sec:eval}

We now evaluate our schemes and compare them with prior work over 
the quantum network simulator NetSquid~\cite{netsquid2020}. 



\para{Graph State Generation Protocol.}
We build our protocols on top of the link-layer protocol of~\cite{sigcomm19}, delegated to continuously generate \epss on a link at a desired rate. 
The key aspect of our generation protocol is that a fusion operation 
is done only when {\it both} the subgraph states (corresponding to the fusion operands) 
have been generated.
On {\bf success} of a fusion operation, the fusion node 
transmits classical information to the terminal nodes of 
its sub-states to manipulate the gate operations on their qubits.
On fusion {\bf failure}, all the qubits for this graph state will be discarded, allowing the protocols in the lower level to generate new link 
\epss and subgraphs.

\begin{figure*}[h]
    \centering
    \includegraphics[width=\textwidth]{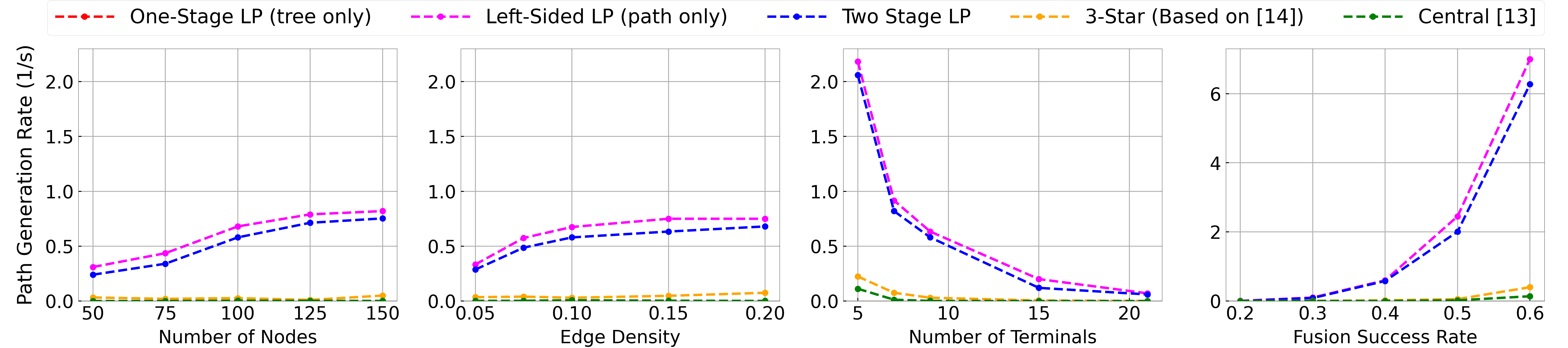}
        \caption{Generation Rates for {\bf path} graph states for various schemes, for varying parameter values, from NetSquid simulations. 
        \blue{The schemes \ghzt and \central have rates of 0.037-0.0018 and 0.0008-0.002 respectively in (a), rates of 0.075-0.0035 and 0.002-0.001  in (b), rates of 0.224-0.0005 and 0.112-0  in (c), and rates of 0.4-0.0001 and 0.15-0  in (d).}
        (\oslp takes exorbitant computation time for a 100-node network. Thus, we plot \llp LP scheme; \rlp LP scheme has similar performance and not shown.)}
    \label{fig:path-ns}
\end{figure*}

\begin{figure*}
    \centering
    \includegraphics[width=\textwidth]{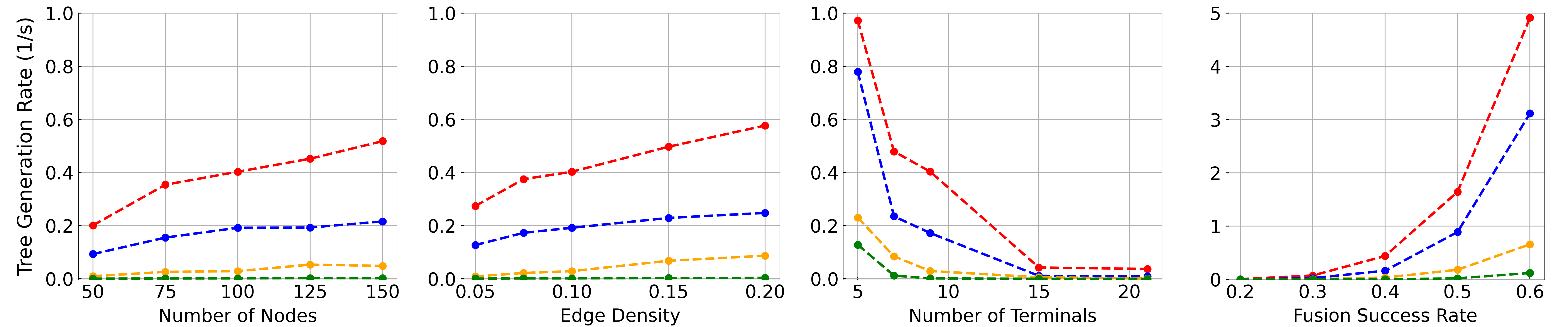}
        \caption{Generation Rates for {\bf tree} graph states for various schemes, from NetSquid simulations. In (c), for 15-21 terminals, \oslp has a generation rate of 0.04-0.03, \tslp have rates of 0.009-0.0007, while \ghzt and \central have rates of 0.005-0.0005 and $10^{{-}6}$-$10^{{-}8}$ respectively.}
    \label{fig:tree-ns}
\end{figure*}

\para{Simulation Setting.}
We generate random quantum networks in a similar way as in 
the recent works~\cite{swapping-tqe-22, sigcomm20}.
By default, we use a network spread over an area of $100 km \times 100 km$.
We use the Waxman model~\cite{waxman}, used to create Internet topologies,
distribute the nodes, and create links. 
We select the terminal nodes that store the graph state within the 
network graph, randomly. 
The path graph state and tree graph state have the same parameter settings. 
We vary the number of nodes from 50 to 150 (default being 100) and 
the number of terminals (i.e., size of the graph state) from 5 to 21 
(default being 9). 
The tree state is as follows: root has 2 children, root's children has 3 children each,
and finally, root's grandchildren have 0-3 children each---yielding a tree of size 9 to 21. 
We vary the edge density from 0.05 to 0.2 with a default value of 0.1.
Each data point is for a 100-second duration simulation in NetSquid.

\softpara{Parameter Values.}
We use parameter values similar to the ones used in~\cite{caleffi, swapping-tqe-22}.
In particular, we use fusion probability of success (\fp) to be 0.4 and latency (\ft) to be 10 $\mu$ secs; 
in some plots, we vary $\fp$ from 0.2 to 0.6.
The atomic-BSM probability of success (\bp) and latency (\bt) always equal their fusion counterparts \fp and \ft. 
The optical-BSM probability of success (\php) is half of \bp. 
For generating link-level \epss, we use atom-photon generation times (\gt) and probability of success (\gp) as 50 $\mu$sec and 0.33, respectively. 
Finally, we use photon transmission success probability 
as $e^{-d/(2L)}$~\cite{caleffi} where $L$ is the channel attenuation length
(chosen as 20km for optical fiber) and $d$ is the distance between the nodes.
As in~\cite{swapping-tqe-22,ghz-qce-23}, we choose a decoherence 
time of two seconds 
based on achievable
values with single-atom memory platforms~\cite{loock20}; 
note that decoherence times of even several 
minutes~\cite{dec-13,dec-14} to hours~\cite{dec-15,ionqmemory21} has been 
demonstrated for other memory platforms. 
\blue{In NetSquid, the storage noise, channel noise and gate noise are modeled using two parameters: the depolarization rate (for decoherence) and the dephasing rate (for operation-driven)~\cite{netsquid2020}. We choose a depolarization rate of $0.02$ and a dephasing rate of $1000$ for our experiments.}

\para{Prior Algorithms Compared.}
For comparison with prior work, we implement
two schemes for EP and one scheme for GHZ: a recent \ghzt-based scheme from~\cite{korean23} (we adapt it for a quantum network) and the flow-based approach (called \central, here) from~\cite{maxflowQCE21}. We describe these below. 
The \ghzt approach~\cite{korean23} is a three-step graph-theoretic scheme: 
simplify the graph state, decompose the simplified graph into star graphs, and 
replace each 
star graph into multiple \ghzt states and determine the order of fusion operations; finally, iterate over the above steps and select the best one.
To adapt the scheme to generate graph states in a quantum network, we generate the required \ghzt state locally in a central node, distribute (via teleportation) the qubits of the \ghzt states 
appropriately, and then fuse them to generate the distributed target graph state. 
The \central approach works by first generating the target graph state 
locally (at an exhaustively picked optimal central node) and then teleporting the qubits of the graph state to the desired terminals.
To continuously generate the graph states at an optimal generation rate, the generation of \epss between $C$ and the terminals is done continuously in parallel with other steps (similar to the max-flow-based approach from~\cite{maxflowQCE21}). 

\begin{figure}
    \centering
    \includegraphics[width=0.8\linewidth]{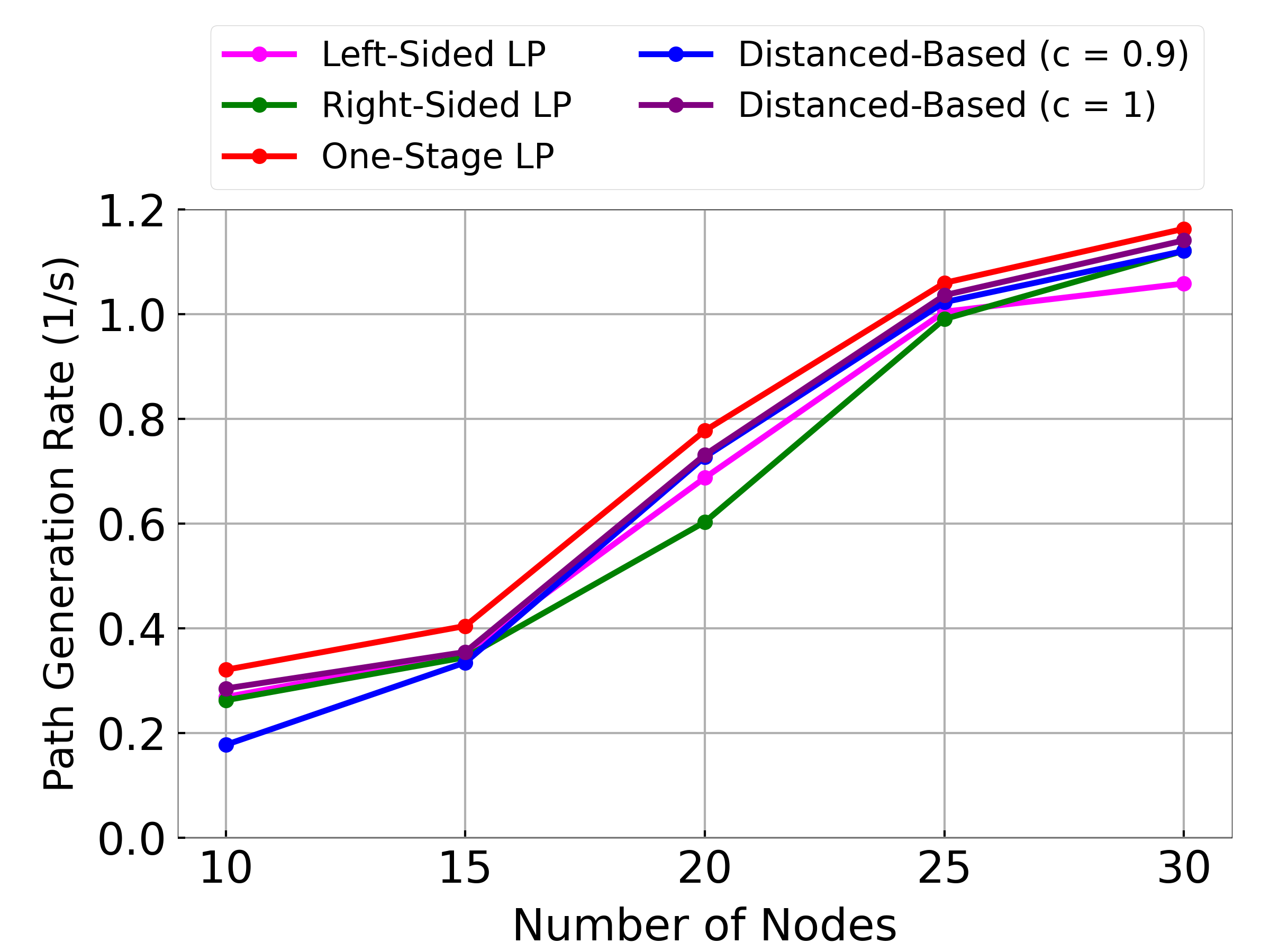}
    \caption{Performance of computationally-efficient approximation schemes for path graph states relative to the optimal \oslp scheme for small network sizes.}
    \label{fig:approx} 
\end{figure}

\para{Our Algorithms.}
For tree graph states, we implement the \oslp and \tslp schemes. For the path states,
the optimal \oslp scheme (\S\ref{sec:opt-path}) takes an exorbitant computation time even for moderate-sized networks; e.g., for a network of 50 (100) nodes, its LP takes 5 (estimated, 120) hours. The \distance approximation schemes perform similarly to \oslp for $c > 0.8$, but also 
incur very high computation time. In contrast, the \llp and \rlp LP schemes take only a few minutes, even for 100-node networks, and perform close to the optimal \oslp scheme. See Fig.~\ref{fig:approx}.
Based on these observations, we only consider \llp and \tslp for path graph states for our further evaluations, which are done over large (default 100 nodes) networks; \rlp performs similarly to \llp and, thus, is not shown.

\begin{figure}
    \centering
     \includegraphics[width=0.5\textwidth]{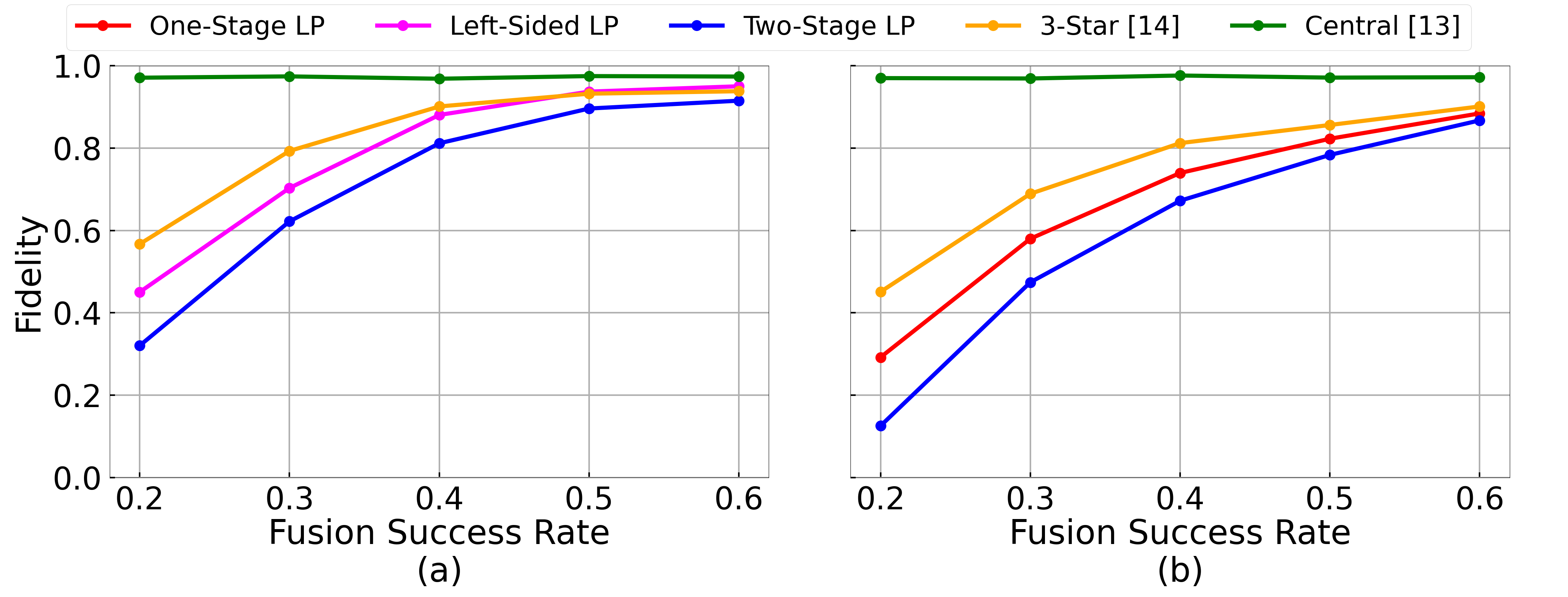}
    \caption{Fidelity of generated graph states. (a) Paths, (b) Trees.}
    \label{fig:fidelity}
\end{figure}

\para{Evaluation Results.} We now present the evaluation results comparing prior work with our techniques.
Figs.\ref{fig:path-ns}-\ref{fig:tree-ns} show the generation rates of various schemes for path and tree graph states, as determined by the 
NetSquid simulations of at least 100-second duration.
In particular, we vary one parameter at a time while keeping the other parameters constant (to their default values). 
We observe that our schemes outperform the \central and \ghzt schemes for both 
path and tree graph states by up to orders of magnitude; in particular, 
the performance gap is about {\bf 100} ($\mathbf{10^6}$) {\bf times} wrt \ghzt (\central) for both path and tree graph states of 21 terminals.
Among our schemes, as expected, \oslp outperforms the \tslp scheme, sometimes with a smaller margin. 
Finally, Fig.~\ref{fig:fidelity}(a)-(b) show the fidelities of the graph states generated in NetSquid simulations by the 
various schemes for both path and tree graph states for varying fusion success rates. We see that
the fidelity of the generated graph states is consistently high. Note that the \central approach has high fidelity since we defer generation of the graph state to {\it after} all the EPs required for teleportation have been successfully generated.
We observe that generation rates in the NetSquid simulations show a similar trend
as those output by the LP solutions (not shown), but the NetSquid simulation rates
were consistently higher; this is because the 2/3 factor estimation in Eqn.~\ref{eqn:tree-rate} only holds when the operand generation rates are equal---this holds in the LP solution but may not hold at higher levels of the level-based structure in an actual simulation.

\cbb
\softpara{Scalability; Runtime.} 
Let $n$ be the number of nodes in the given quantum network, and $m$ be the size (number of terminals) of the target graph state. The complexity of the number of hypervertices and hyperedges in the hypergraph for the LPs used by the various schemes is as follows. (Note that the number of {\em variables} in the LP is equal to the number of hyperedges in the corresponding hypergraph). We also give actual numbers for the specific case of the default setting of 100 nodes and 9 terminals. 
\begin{itemize}
    \item \oslp scheme for paths results in $O(n^2 m^2)$ hypervertices and $O(n^3 m^4)$ hyperedges. For the default setting, there were 25,808 hypervertices and 814,475 hyperedges.
    \item \tslp scheme for trees results in $O(n^2 + m^3)$ hypervertices and $O(n^3 + m^4)$ hyperedges. For the default setting, there were 10,444 hypervertices and 491,105 hyperedges.
    \item \oslp scheme for trees results in $O(n^2 + n m^3)$ hypervertices and $O(n^3 + n m^4 + n^2 m^3)$ hyperedges. For the default setting, there were 14,007 hypervertices and 634,970 hyperedges.
\end{itemize}

\blue{In our evaluations, our presented schemes take 1-3 minutes to run for our default settings in 100-node networks on a 5GHz machine (see Table~\ref{tab:runtime}). As expected from the above complexity and numbers, our schemes for the generation of tree graph states take less time than those for path graph states.}
Overall, the runtime observed is tolerable overhead, especially for the case of continous generation of graph
states as the overhead to determine an efficient generation {\em scheme} is only one-time. Note that 
optimizing generation latency of graph states also 
minimizes their fidelity degradation due to minimal storage time during generation.
\cbl




\begin{table}[h]
    \centering
    \caption{\blue{Average Runtime for Different Schemes in 100-node quantum networks with default settings.)}}
    \begin{tabular}{lcc}
        \toprule
        \textbf{Scheme} & \textbf{Runtime (seconds)} \\
        \midrule
        Tree Graph State Two-Stage & 54  \\
        Tree Graph State One-Stage & 90  \\
        Path Graph State Two-Stage & 82  \\
        Path Graph State Left-Sided One-Stage & 126  \\
        \bottomrule
    \end{tabular}
    \label{tab:runtime}
\end{table}

\section{\bf Conclusions}
\label{sec:conc}

We have developed a framework for developing optimized generation and distribution of classes of multipartite graph states under appropriate constraints while considering the stochasticity of the underlying processes.   
\blue{Our methods can also be used to improve the generation rates of given target graph
states using desired operations and potential intermediate states (e.g.,~\cite{epping2016large,pirker2019quantum}) by formulating an appropriate LP with the  potential intermediate states.}
Our future work is focused on developing provably optimal generation schemes under fewer assumptions and/or for other useful classes of graph states.



\section*{Acknowledgment}
This work was supported by NSF awards FET-2106447 and CNS-2128187.
\bibliographystyle{IEEEtran}
\bibliography{all_shorten}
\begin{IEEEbiography}[{\includegraphics[width=1.1in,height=1.3in,clip,keepaspectratio]{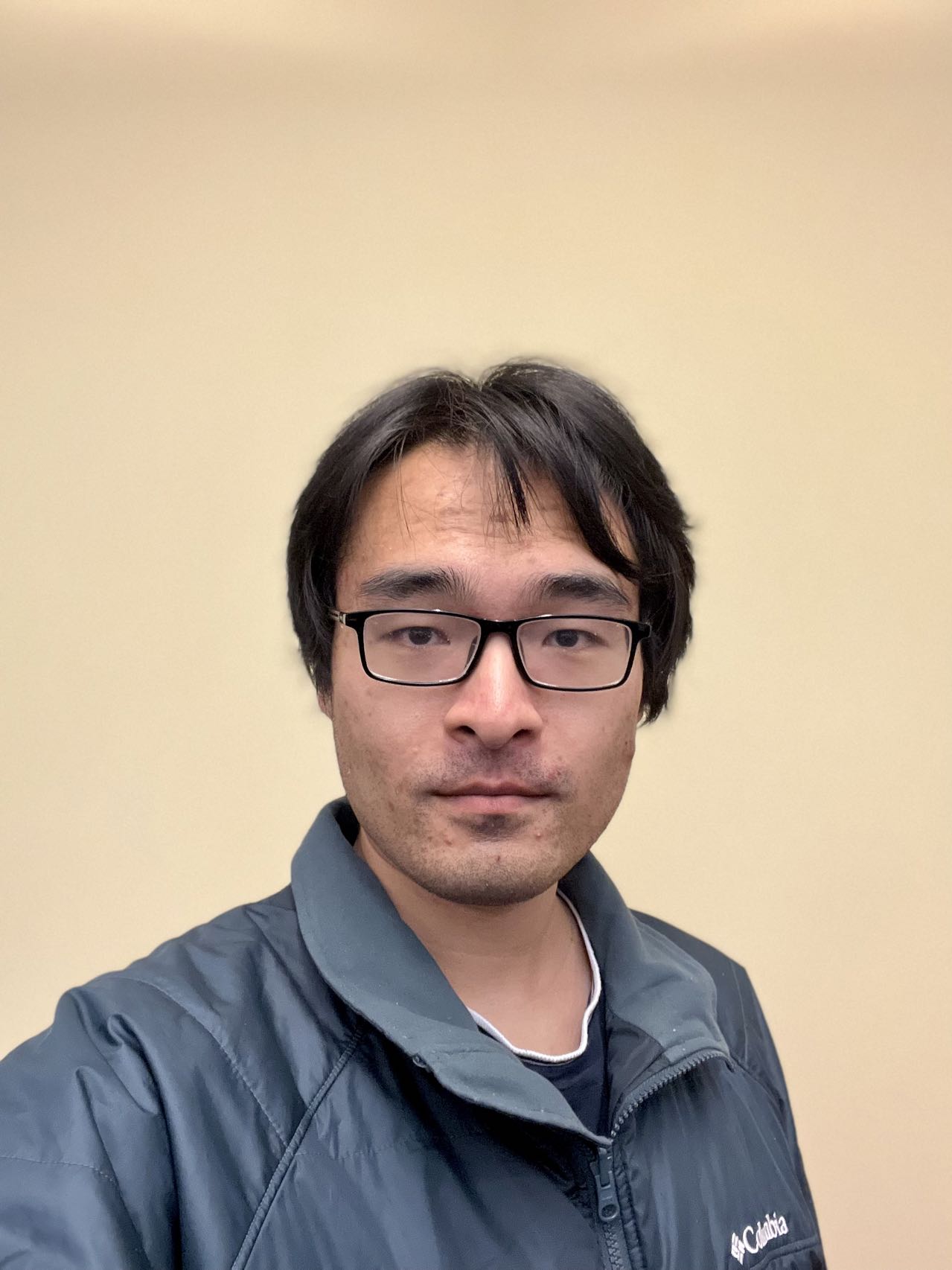}}]{Xiaojie Fan} 
received his B.S. degree in spatial information and digital technology from China University of Geosciences in Wuhan, China in 2020 and the M.S. degree in computer science in Huazhong University of Science and Technology in Wuhan, China in 2022. He then joined the PhD program in Stony Brook University at the Department of Computer Science. His research interests lies in quantum networks.
\end{IEEEbiography}

\begin{IEEEbiography}[{\includegraphics[width=1.1in,height=1.3in,clip,keepaspectratio]{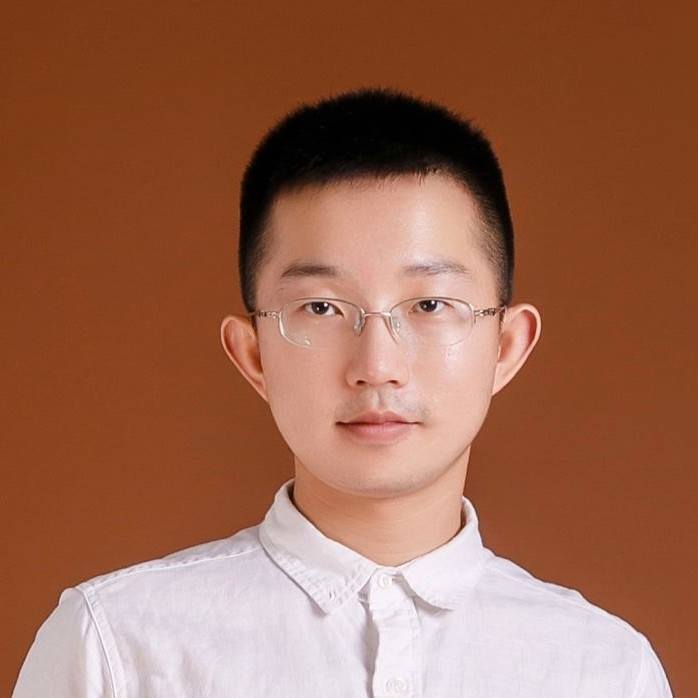}}]%
{Caitao Zhan}
received his B.S. degree in computer science and technology from China University of Geosciences in Wuhan, China in 2017.
He received his PhD in computer science from the Department of Computer Science at Stony Brook University in 2024.
He is currently a Postdoc researcher at Argonne National Laboratory.
He does research in the broad area of computer networks. 
His current research focus is quantum communication networks, quantum computing, and quantum sensing.
\end{IEEEbiography}

\begin{IEEEbiography}[{\includegraphics[width=1.1in,height=1.3in,clip,keepaspectratio]{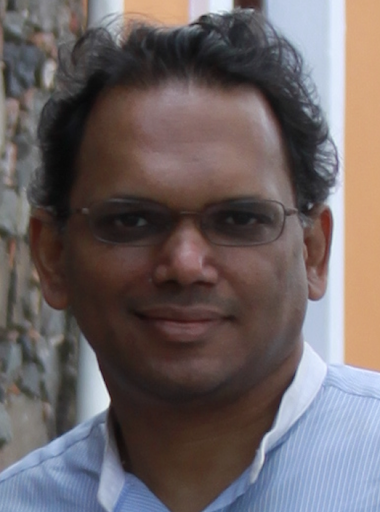}}] {Himanshu Gupta}
is a professor of computer science at Stony Brook University, where he has been a faculty since 2002. His area of research has been in wireless networks, with recent focus on free-space optical communication networks and spectrum management. His current research focuses on quantum networks and communication, and distributed quantum algorithms. He graduated with an M.S. and Ph.D. in Computer Science from Stanford University in 1999, and a B.Tech. in Computer Science and Engineering from IIT Bombay in 1992.
\end{IEEEbiography}


\begin{IEEEbiography}[{\includegraphics[width=1.1in,height=1.3in,clip,keepaspectratio]{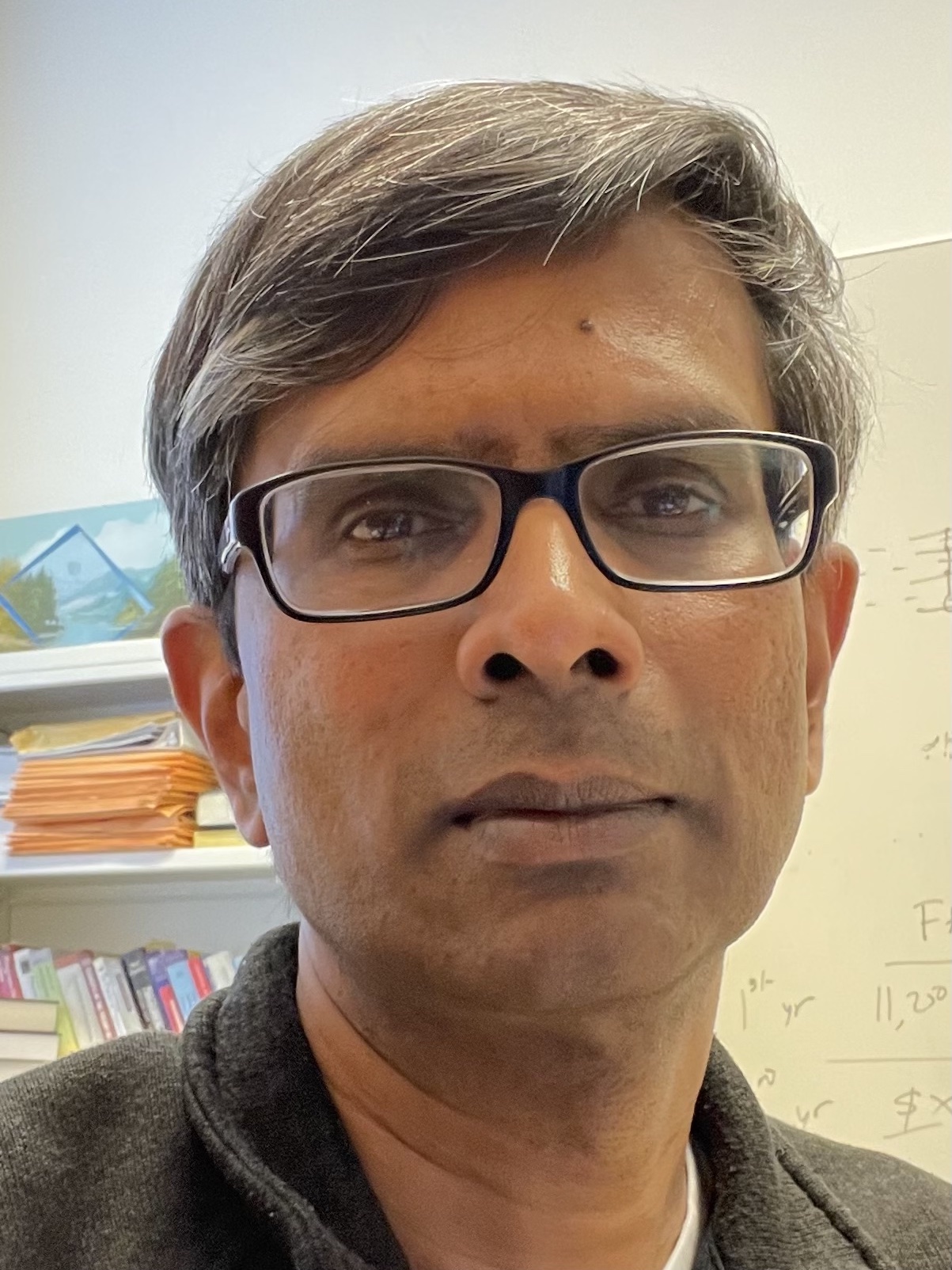}}] {C.\ R.\ Ramakrishnan}
is a professor of computer science at Stony Brook University, where he has been a faculty since 1997. His area of research has been in logic programming and verification, with recent focus on analyzing properties of probabilistic programming. His current research focuses on quantum networks and communication, and distributed quantum algorithms. He graduated with an Ph.D. in Computer Science from Stony Brook University in 1995, and an M.Sc. (Tech.) in Computer Science from BITS Pilani in 1987.
\end{IEEEbiography}

\EOD

\end{document}